\documentclass[preprint2]{aastex}

\def\boldalpha{\mbox{\boldmath$\alpha$}}
\def\boldbeta{\mbox{\boldmath$\beta$}}
\def\hboldalpha{\mbox{\boldmath$\hat\alpha$}}
\def\hboldbeta{\mbox{\boldmath$\hat\beta$}}
\def\bolddelta{\mbox{\boldmath$\delta$}}
\def\boldgamma{\mbox{\boldmath$\gamma$}}
\def\boldtheta{\mbox{\boldmath$\theta$}}
\def\boldeta{\mbox{\boldmath$\eta$}}
\def\boldpsi{\mbox{\boldmath$\psi$}}
\def\boldb{\mbox{\boldmath$b$}}
\def\bolde{\mbox{\boldmath$e$}}

\def\boldx{\mbox{\boldmath$x$}}
\def\glfit{\texttt{GLFit}}
\def\simgeq{\raisebox{-0.8ex}{$\stackrel{\textstyle >}{\sim}$}}
\def\simleq{\raisebox{-0.8ex}{$\stackrel{\textstyle <}{\sim}$}}
\def\eqq#1{Equation~(\ref{#1})}

\begin{document}
\slugcomment{CVS $Revision: 1.37 $ $Date: 2006/06/29 15:12:11 $}

\title{Shear Recovery Accuracy in Weak Lensing Analysis with \\
Elliptical Gauss-Laguerre Method}

\author{Reiko Nakajima and Gary Bernstein}
\affil{Department of Physics \& Astronomy, University of Pennsylvania, 
209 S.\ 33rd St., Philadelphia, PA 19104}

\begin{abstract}
We implement the Elliptical Gauss-Laguerre (EGL) galaxy-shape
measurement method  
proposed by Bernstein \& Jarvis (2002) [\citet{BJ02}]
and quantify the shear recovery accuracy in weak lensing analysis.
This method uses a deconvolution fitting scheme to remove the 
effects of the point-spread function (PSF).
The test simulates $>10^7$ noisy galaxy images convolved with 
anisotropic PSFs, and attempts to recover an input shear.
The tests are designed to be immune to shape noise, selection biases,
and crowding.
The systematic error in shear recovery is divided into two classes,
calibration (multiplicative) and additive, with the latter 
arising from PSF anisotropy.  At S/N $>50$, the deconvolution 
method measures the galaxy shape and input shear to $\sim1\%$
multiplicative accuracy, 
and suppresses $>99\%$ of the PSF anisotropy.
These systematic errors increase to $\sim4\%$ for the worst 
conditions, with poorly resolved galaxies at S/N $\simeq20$.
The EGL weak lensing analysis has the best demonstrated accuracy
to date, sufficient for the next generation of weak lensing surveys.

\end{abstract}


\section{Introduction}

Weak gravitational lensing,
the shearing of galaxy images by gravitational bending of light,
is an effective tool to probe the large-scale matter distribution 
of the universe.
It is also a means to measure the cosmological parameters 
by comparing observation to numerical simulations of large scale 
structure growth~\citep{bartelmann01}.
There are many weak lensing (WL) surveys underway to obtain 
the cosmological parameters to higher precision, and in particular
to probe the evolution of the dark energy by observing its effects 
on the evolution of matter distribution
(DLS\footnote{Deep Lens Survey: {\tt dls.bell-labs.com/}}, 
CFHTLS\footnote{Canada France Hawaii Telescope Legacy Survey: 
{\tt www.cfht.hawaii.edu/Science/CFHLS/}}).

The WL signal is very subtle, however; it is necessary to measure
these small distortions  (typical shear $\gamma\sim 1\%$) in the
presence of optical distortions and the asymmetric 
point-spread-function (PSF) of real-life imaging.  
The level of systematic error in the WL measurement methods are currently 
above the statistical accuracy expected from future wide and deep WL 
surveys
(Pan-STARRS\footnote{Panoramic Survey Telescope \& Rapid Response System:
{\tt pan-starrs.ifa.hawaii.edu/}}, 
SNAP\footnote{Supernova / Acceleration Probe: {\tt snap.lbl.gov/}}, 
LSST\footnote{Large Synoptic Survey Telescope: \tt www.lsst.org/}, 
SKA\footnote{Square Kilometre Array: \tt www.skatelescope.org/}).
Because there are no ``standard shear'' lenses on the sky,
shear-measurement techniques are tested by applying them to artificial
galaxy images and seeing if one can correctly extract a shear applied
to the simulation.  In most cases, the recovered shear can be written
as $\gamma_{\rm out} = m \gamma_{\rm in} + c$.  Departures from the
ideal $m=1$ we will term ``calibration'' or ``multiplicative'' errors
and quote as percentages.  Deviations from the ideal $c=0$ can result
from uncorrected asymmetries in the PSF and optics, and will be termed
``additive errors'' or ``incomplete PSF suppression.''

Such tests of the most widely applied analysis method
\citep{KSB}[KSB], find $m=$0.8--0.9, but this coefficient is
implementation dependent~\citep{erben01,bacon01}, and depends upon the
characteristics of the simulated galaxies.
Hirata \& Seljak (2003) [\citet{Hirata03}] 
demonstrate that various PSF-correction methods can
produce shear measurements miscalibrated by a few \% to 20\% or
more.  Heymans et al.\@ (2005) [Shear TEsting Programme, (\citet{STEP})] 
present testing of many existing shear-measurement pipelines using a 
common ensemble of sheared simulated images.  
These methods show a median calibration
error of 7\%, although some (the BJ02 rounding kernel method, 
an implementation of a KSB method, as well as the one described 
in this paper) show no calibration error, 
to within the $\sigma_m\approx1\%$ noise level of the first STEP tests.
Although the statistical accuracy in past surveys was comparable to the
7\% systematics, it is expected to be well below 1\% in future surveys.
Hence, understanding and eliminating the WL systematic errors 
require the most urgent attention today.

In this paper, we implement the elliptical Gauss-Laguerre (EGL)
deconvolution method as described in BJ02, 
and subject it to a series of tests designed to be 
more stringent than any previous test of WL measurements.
The deconvolution method is distinct from the \citet{Jarvis03}
method, also described BJ02, in which the anisotropic PSF effects are 
removed using a ``rounding kernel'' instead.

WL testing regimes are of two types: in end-to-end tests ({\it e.g.}
STEP), one produces simulated sky images with a full population of
stars and galaxies, analyzes them with the
same pipeline as one would real data, then checks the output shear for
veracity.  We perform here more of a dissection, in which we analyze
the performance of the method one galaxy type at a time, and vary the
parameters of the galaxy and PSF images to determine which, if any,
conditions cause the measurement to fail.  While lacking the realism
of an end-to-end test, this allows us to isolate and fix weaknesses.
If we can demonstrate that the method succeeds under a set of
conditions that will circumscribe those found on the real sky, then we
can have confidence that our method is reliable, whereas end-to-end
testing is reliable only to the extent that the simulated sky
reproduces the characteristics of the real sky.

We investigate here the performance of our EGL method across the
range of noise levels, degree of resolution by the PSF, pixel sampling
rates, galaxy ellipticity, and PSF ellipticity, using both highly
symmetric and asymmetric galaxy shapes.  We test not only the accuracy
of shear recovery, but also the accuracy of the shear uncertainty
estimates. 

The EGL method is further elaborated in \S2, while the 
implementation, \glfit, is detailed in \S3.
The shear accuracy test procedure is described in \S4.  The conditions
under which the shape measurement succeeds, and the accuracy of its
estimates of shear, are presented in \S5.  
Previous dissection tests include \citet{Hirata03} and \citet{Kuijken}.
The former studies the performance of several methodologies on varied 
galaxy and PSF shapes/sizes in the absence of noise.  The latter study 
verified its ``polar shapelet'' method to better than 1\% calibration 
accuracy.  In \S6 and \S7 we conclude with comparisons to other
shape-measurement methodologies and tests, and draw inferences for
future surveys.

\section{Overview of Shape Measurement}

The task of this weak lensing methodology is to assign some {\em shape}
$\bolde_i$ to observed galaxy $i$, then to derive from the ensemble
$\{\bolde_i\}$ an estimate of the applied lensing {\em shear}
\boldgamma.  More precisely, a shape analysis can only determine the
{\em reduced shear} $g=\gamma/(1-\kappa)$, where $\kappa$ is the lens
convergence. Following BJ02, we use distortion \bolddelta\ to describe
the shear, where $\delta = 2g/(1+g^2)$ ($\delta\simeq2g$ for $g\ll1$).
In this paper, both the shear and the shapes are expressed as distortions;
while in other WL literatures, shear is usually expressed as $g$.

\subsection{Defining Shapes using Shear}
\label{shape-as-shear}

Following BJ02, we will quantify the lensing by
decomposing its magnification matrix ${\bf M}$ into a diagonal
dilation matrix ${\bf D}_\mu=e^\mu{\bf I}$ and a unit-determinant
symmetric shear matrix ${\bf S}_{\boldeta}$:
\begin{eqnarray}
{\bf M} & = & {\bf D_\mu} {\bf S}_{\boldeta}
 = {\bf D_\mu} {\bf R}_\beta  {\bf S}_\eta {\bf R}_{-\beta } \\
{\bf R}_\beta & = & \left(\begin{array}{cc}
  \cos \beta & -\sin \beta \\
  \sin\beta & \cos \beta
\end{array}\right) \\
{\bf S}_\eta & = & \left(\begin{array}{cc}
  e^{\eta/2} & 0 \\
  0 & e^{-\eta/2}
\end{array}\right) \\
\boldeta & \equiv &(\eta_+, \eta_\times) = (\eta \cos 2\beta, \eta \sin 2\beta)
\end{eqnarray}
where $\beta$ is the direction of the shear axis, and 
$\eta$ is a measure of shear.
The ``conformal shear'' $\eta$ can be reparameterized as the
distortion $\delta=\tanh\eta$ or the reduced shear
$g=\tanh\eta/2$.  

\begin{figure}[!tbp]
\plotone{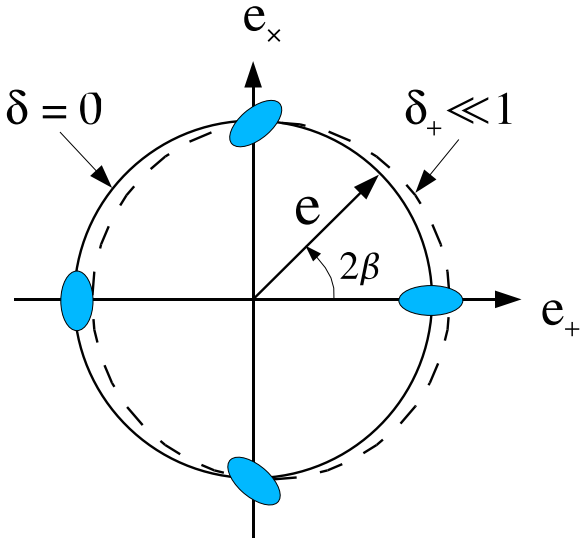}
\figcaption{The ring test.  
In the ring test, objects that have ellipticity 
magnitude $e=\sqrt{e_+^2 + e_\times^2}$ are distributed evenly in $\beta$ 
on a ring in the ellipticity plane (solid circle).  
In the absence of shear $\delta$, 
the average of the measured shapes should be zero.
In the presence of small shear $\delta_+$, the distribution 
of objects in the ellipticity plane are displaced such that they are
no longer arranged with their center on the origin.  The average of 
all the shapes on the displaced ring should behave as
Eq.~(\ref{avgweakmap}), to the first order in $\delta_+$.
\label{fig:ringtest}}
\end{figure}

The shape must be assigned to an image of a galaxy with
some surface-brightness distribution $I(\boldtheta)$.  Initially we
will ignore the effects of PSF convolution on the observed image.  The
BJ02 definition of shape is to specify
roundness criteria or ``circularity tests,''
$M_+(I)$ and $M_\times(I)$, that operate on $I$ to yield one scalar for
each component of the shear---typically these are quadrupole moments.
The object is deemed circular 
($\bolde=0$) if $M_+(I)=M_\times(I)=0$.  If the object is not
circular, then we assign to the object the shape \boldeta\ which
yields the solutions 
\begin{equation}
\label{shapedef}
M_+[I({\bf S}_{\boldeta}\boldtheta)] = 
M_\times[I({\bf S}_{\boldeta}\boldtheta)] = 0,
\end{equation}
{\it i.e.} we find the shear \boldeta\ that, when applied to the 
coordinate system \boldtheta, makes the image appear circular in 
that coordinate system, and declare the galaxy shape to be this shear.
Any circularity test will do, as long as \eqq{shapedef} has a unique 
solution \boldeta; in 
particular the matrix $d{\bf M} / d\boldeta$ must be non-singular.
The shape \bolde\ is defined by $e=\tanh\eta$, keeping the position
angle $\beta$.
Defining shape in this way with
a suitable circularity test has the virtue that the 
effect of a lensing distortion \bolddelta\ upon the galaxy shape
\bolde\ is completely defined by the multiplication of
shear matrices.  In particular, the component-wise formulae for
transformation of a shape under a shear must take the form given by
\citet{escude91}: 
\begin{mathletters}
\label{emap}
\begin{equation}
e^\prime_+ = { {e_+ + \delta_+ +
(\delta_\times/\delta^2) [ 1 - \sqrt{1-\delta^2} ] 
(e_\times\delta_+ - e_+\delta_\times) }
\over { 1 + \bolde \cdot \bolddelta} };
\end{equation}
\begin{equation}
e^\prime_\times = { {e_\times + \delta_\times +
(\delta_+/\delta^2) [ 1 - \sqrt{1-\delta^2} ] 
(e_+\delta_\times - e_\times\delta_+) }
\over { 1 + \bolde \cdot \bolddelta} }.
\end{equation}
\end{mathletters}
We can take the limit of a weak shear $\delta_+\ll 1$,
$\delta_\times =0$:
\begin{mathletters}
\label{weakmap}
\begin{eqnarray}
\label{weakmap1}
e_+^\prime & = & e_+ + (1-e_+^2)\delta_+ \\
e_\times^\prime & = & e_\times  - e_+e_\times\delta_+
\end{eqnarray}
\end{mathletters}
BJ02 describe (\S5) a scheme for optimally weighting and combining an
ensemble of shapes to produce an accurate estimate of the distortion
\bolddelta. This scheme is predicated on the assignment of shapes that
transform under shear according to \eqq{emap}.  Hence to test the
accuracy of our methodology in recovering weak lensing shear, {\em we
  need only test that assigned shapes transform under shear as in
Equations~(\ref{weakmap}).} Furthermore, the isotropy of the Universe
guarantees 
that the $\bolde_i$ of an unlensed population will be uniformly distributed in
$\beta$, hence we need only verify that Equations~(\ref{weakmap}) hold
when averaged over an ensemble of galaxies with fixed unlensed $|e|$ but random
orientations:
\begin{mathletters}
\label{avgweakmap}
\begin{eqnarray}
\langle e_+^\prime \rangle & = & \langle e_+ \rangle + 
\langle 1-e_+^2 \rangle \delta_+ = (1-e^2/2) \delta_+ \\
\langle e_\times^\prime \rangle & = & 
\langle e_+ \rangle + \langle e_+e_\times \rangle \delta_+ = 0
\end{eqnarray}
\end{mathletters}
Here the brackets refer to averaging over the pre-lensing orientation
$\beta$. The second order term in $\delta_+$ vanishes, so these
equations are valid to $O(\delta^3)$.
We refer to this as the {\em ring test} (Fig.~\ref{fig:ringtest}), 
since we construct an
ensemble of test galaxies which form a ring in the \bolde\ plane, then
shear them, measure their shapes, and take the mean.  We also note
that as a special case, we should obtain $\langle
\bolde^\prime\rangle=0$ when there 
is {\em no} applied shear.  When there is no PSF or the PSF is
symmetric under 90\arcdeg\ rotation, then this result holds for any
measurement scheme that is symmetric under inversion or exchange of
the $x$ and $y$ axes of images.  But for an asymmetric PSF, this is a
stringent test of the ability of the shape-measurement technique to
remove the effects of the PSF from the galaxy shapes.

\subsection{Fitting to Basis Functions}
\label{fits}

The circularity test described in BJ02 involves decomposing the
pre-convolution surface brightness distribution of the galaxy,
$I(\boldtheta)$, into the Gauss-Laguerre set of orthonormal basis
functions in the plane.  We consider here a general set of 
two dimensional functions
$\{\psi_i(\boldtheta)\}$ that are complete (though not necessarily
orthogonal) over the plane.  Any set of complete functions can be
transformed to a new complete set $\{\psi^E_i\}$ via
\begin{equation}
\psi_i^E(\boldtheta) \equiv \psi_i({\bf E}^{-1}\boldtheta).
\end{equation}
Here ${\bf E}$ is any remapping of the sky plane that compounds a
displacement $\boldx_0$, a shear $\boldeta$, and a dilation
$\mu$.  We refer to ${\bf E}$ as the 
{\em basis ellipse} of the function set: if the unit circle is an
isophote of $\psi_i$, then $\{\boldx_0,\mu,\boldeta\}$ describe the
center, size, and shape of an ellipse that traces the same isophote
for $\psi^E_i$.  We will also use ${\bf E}$ to refer to this set of 5
parameters that defines an ellipse; it will be clear from context
whether we are referring to the parameter vector or the coordinate
transform. 
For any {\bf E} we must have
\begin{equation}
\label{ibpsi}
I(\boldtheta) = \sum_i b_i \psi_i^E(\boldtheta) = \boldb \cdot
\boldpsi^E(\boldtheta),
\end{equation}
for (at least) one vector \boldb\ of coefficients.  For the function
set to be complete, the vectors must be infinite-dimensional, but 
a real measurement will model $I$ with a finite subset of the basis
functions.  The model must be fit to the observed image plane which we
denote with coordinates \boldx.  

The action of the atmosphere, optics, and detector will operate on
$I(\boldtheta)$ to produce an observed surface-brightness distribution 
$I_o(\boldx)$ on the detector plane.  This observation operator
$C$ likely includes convolution by the PSF and perhaps some
distortion by the optics.  We will assume that this operation is known
and that it is linear over surface brightness, so that $C(aI_1 +
bI_2) = aC(I_1) + bC(I_2)$, where $a$ and $b$ are any two scalars.
In this case, \eqq{ibpsi} must also imply that
\begin{eqnarray}
\label{ibphi}
I_o(\boldx) & = & \sum_i b_i \phi_i^E(\boldx), \\
\phi^E_i(\boldx) & \equiv & C\left[\psi^E_i(\boldtheta)\right].
\end{eqnarray}

The observed-plane brightness is sampled at the centers of the pixels
$\boldx_p$ yielding measurements $I_p$ with uncertainties $\sigma_p$,
assumed henceforth to be Gaussian.  The model
which maximizes the likelihood of the decompositions
(\ref{ibpsi}) and
(\ref{ibphi}), given the data is that which minimizes
\begin{equation}
\label{chisq}
\chi^2 = \sum_{p\in\,{\rm pixels}} {
\left[ I_p - \sum_i b_i \phi_i^E(\boldx_p)\right]^2
\over \sigma_p^2}
\end{equation}
We call this a ``forward fit'' to the galaxy image, because we are
positing a distribution $\boldb\cdot\boldpsi^E(\boldtheta)$ of flux on
the sky, then propagating this to the detector plane, where we compare
the model to the observations.  Our task will be to find the ${\bf E}$
for which the $\chi^2$-minimizing vector \boldb\ satisfies the
circularity constraints.\footnote{
Note this condition is subtly different from that imposed by
\citet{Kuijken}: his ellipticity is that which minimizes $\chi^2$ if
the \boldb\ is always constrained to be circular.  We do not yet
understand the significance, if any, of this distinction.}

For fixed ${\bf E}$, the $\chi^2$ minimization is a linear
least-squares problem over
\boldb\ with the usual analytic solution
\begin{eqnarray}
\label{sumpix}
\boldb & = & \boldalpha^{-1} \cdot \boldbeta \\
\beta_i & \equiv & \sum_p I_p \phi_i^E(\boldx_p)/ \sigma_p^2 \\
\alpha_{ij} & \equiv & \sum_p \phi_i^E(\boldx_p)
               \phi_j^E(\boldx_p) / \sigma_p^2 .
\end{eqnarray}
If the $\psi_i$ are orthogonal {\em and} the noise is stationary and
white {\em and} the sampling approaches the continuum limit {\em and}
the PSF approaches a delta-function, then
\boldalpha\ will be diagonal (modulo some complex conjugation
operations) and the solution is very simple.  These conditions are
not, however, typically met by real data---in particular the sampling
and PSF conditions---so it is not safe to use
orthogonality to decompose the observed data \citep{BJ02, Massey, BHW}. 

For finite data, the solution must be done over some truncated basis
set.  We will assume that $N$ of the basis functions are being used,
so that \boldalpha, \boldbeta, and \boldb\ have dimension $N$.
The $\chi^2$-minimizing \boldb\ is
uniquely defined, regardless of whether the $\psi_i$ are complete or
orthogonal, as long as $\boldalpha$ is not singular.

To follow the BJ02 prescription we must define circularity tests and
iterate the shear in ${\bf E}$ until the two circularity components vanish.  
At the same time, we may wish to define null tests for the centroid
$\boldx_0$ and/or size $\mu$ of the galaxy, so there will be some
number $K\le5$ such statistics ($K=5$ when there is a null test for 
each of the parameters of ${\bf E}$).  Any
test which is linear over the true intensity $I(\boldtheta)$ may be
expressed as a linear operation on \boldb.  Hence the task of the
fitter is to adjust $K$ components of the ${\bf E}$ parameters until
the $\chi^2$-minimizing \boldb\ satisfies
\begin{equation}
\label{mb0}
{\bf M} \cdot \boldb = {\bf 0}.
\end{equation}
${\bf M}$ is a $K\times N$ matrix that defines our tests for matching
the basis ellipse to the center, size, and shape of the
pre-convolution galaxy.

Once \eqq{mb0} is satisfied for a chosen set of basis functions $\psi_i$ 
and any choice of the circularity-test matrix ${\bf M}$,
we obtain a well-defined measurement of the shape \bolde\ 
(or more generally the defining ellipse {\bf E}) of the 
pre-convolution galaxy image.  Since these shapes are defined by
shear operations on the $\boldtheta$ plane, they should transform
according to \eqq{emap} and be amenable to all the weighting and
responsivity schemes of BJ02.

Before describing how we solve for ${\bf E}$, we summarize the
conceptual and practical advantages of combining a
transformation-based definition of shape with forward fitting of a
pre-convolution model to the observed pixels.  Note the fitting
approach has been advocated by several authors \citep{  FischerProFit, 
BJ02,Massey, Kuijken}.
\begin{itemize}
\item The \bolde\ values determined in this way transform in a
  well-known way under lensing shear.  The response of a galaxy
  population to lensing shear can be calculated without recourse to
  empirical calibration factors beyond the distribution of the
  \bolde\ values themselves.
\item The forward-fitting procedure can in principle work with any
  sort of PSF, even those---such as the Airy function---that have
  divergent second moments.
\item The forward-fitting procedure properly handles pixelization and
  sampling by the detector.  Furthermore any aliasing ambiguities will
  properly propagate into the covariance matrix for \boldb, and as
  described below can be propagated into uncertainties in ${\bf E}$.
\item Pixels rendered useless by defects or cosmic rays are easily
  omitted from the measurement.
\item The method is easily extended to simultaneous fitting of
  multiple images with distinct PSFs.  The sum over pixels in
  \eqq{sumpix} is run over all pixels in all exposures of the galaxy.
  All exposures share the same \boldb, \boldpsi, and ${\bf E}$
  vectors, but have distinct observation operators $C$ and hence
  different $\phi_i$.  Information available in the best-seeing images
  is properly exploited.
\item The method is easily adapted to the analysis of $uv$-plane
  interferometric data rather than image-plane data.
\end{itemize}

\subsection{Iterating the Fit}
The desired ellipse parameters ${\bf E}=\{x_0, y_0, \mu, \eta_+,
\eta_\times\}$ enter the least-squares fit non-linearly, hence some
iterative scheme (or Markov chain, {\it cf.} \citet{Bridle}) is
required to determine the values that meet the constraint
(\ref{mb0}).  For a chosen ${\bf E}$, the determination of \boldb\ has
the rapid solution (\ref{sumpix}).  If the current estimate ${\bf
  E}_0$ yields a $\boldb_0$  that does not meet the circularity
condition, the Newton-Raphson iteration would be
\begin{equation}
\label{deltae}
{\bf E}_0 \rightarrow {\bf E}_0 + \delta{\bf E}, \qquad
\delta{\bf E} = -\left({\bf M} {d\boldb \over d{\bf E}}\right)^{-1}
({\bf M} \boldb_0).
\end{equation}
The derivative $d\boldb / d{\bf E}$ follows from noting the effect of
a small change $\delta{\bf E}$ to the basis of \boldpsi:
\begin{eqnarray*}
\psi^{E+\delta E}_i(\boldtheta) & \approx & \psi_i^E(\delta{\bf E}^{-1}
\boldtheta) \\
 & = & \psi_i^E(\boldtheta) + \sum_k \delta E_k \sum_{j=0}^{\infty}
G_{kij}\psi^E_j(\boldtheta) \\
\Rightarrow \phi^{E+\delta E}_i(\boldx) & \approx 
& \phi^E(\boldx) + \sum_k \delta E_k \sum_{j=0}^{\infty}
G_{kij}\phi^E_j(\boldx).
\end{eqnarray*}
Here ${\bf G}_k$ is the {\em generator} for the transformation
indicated by $k$th parameter of ${\bf E}$---either translation,
dilation, or 
shear.  These matrices are fixed by the choice of basis functions
$\{\psi_i\}$.  These alterations to the basis-function values can be
propagated through the solution (\ref{sumpix}) to give the
perturbation to \boldb:
\begin{eqnarray}
\nonumber\delta\boldbeta & = & \sum_k \delta E_k {\bf G}_k \hboldbeta \\ 
\nonumber\delta\boldalpha & = & \sum_k \delta E_k (
                {\bf G}_k \hboldalpha + \hboldalpha^T {\bf G}_k^T) \\
\nonumber\Rightarrow \ \boldb + \delta\boldb & = & (\boldalpha +
\delta\boldalpha)^{-1} (\boldbeta + \delta\boldbeta) \\
\nonumber & \approx & \boldalpha^{-1}\boldbeta
 + \boldalpha^{-1}(\delta\boldbeta) - \boldalpha^{-1}
\nonumber (\delta\boldalpha) \boldalpha^{-1} \boldbeta \\
\label{dbde1}
\nonumber\Rightarrow \quad
\delta\boldb & = & \sum_k \delta E_k \left[ 
\boldalpha^{-1} {\bf G}_k (\hboldbeta - \hboldalpha \boldb_0) \right.\\
& & \quad\quad\quad\quad \quad \ \ 
- \left.\boldalpha^{-1} \hboldalpha^T {\bf G}_k^T \boldb_0 \right].
\end{eqnarray}
The matrix $d\boldb / d{\bf E}$ is apparent from this last equation.
Here we have taken the generator matrices ${\bf G}_k$ to each be
$N\times\infty$; the vector $\hboldbeta$ now must be, in general,
augmented to infinite dimension, and we also take $\hboldalpha$
to be $\infty\times N$.  Note that the parenthesized
portion of the solution (\ref{dbde1}) would vanish if not for the
distinction between the truncated and infinite-dimensional versions of
\boldalpha\ and \boldbeta, since the first $N$ elements of
$\hboldbeta - \hboldalpha \boldb_0$ are zero.  Likewise we could
set the initial $\boldalpha^{-1} \hboldalpha^T$ of the final term
to identity if not for the truncation, in which case the
transformation would become the very simple $\delta\boldb = ({\bf
  G}^T\boldb_0)\cdot \delta{\bf E}$ (with some abuse of notation
here).  Since we are using $d\boldb/d{\bf E}$ only to help us iterate
the solution for ${\bf E}$, we could use this simple approximation, or
extend to some order beyond $N$ using \eqq{dbde1}.

\subsection{The Gauss-Laguerre Decomposition}
\label{gldecomposition}
We use the Gauss-Laguerre decomposition in our shape measurements.
These are the eigenfunctions of the 2-dimensional quantum harmonic
oscillator, and are most compactly expressed as complex functions
indexed by two integers $p,q \ge 0$:
\begin{equation}
\label{laguerre1}
\psi_{pq}(r,\theta)  =  
	{ {(-1)^q} \over {2\pi}}
	\sqrt{ {q!} \over {p!} }
	\;r^m e^{im\theta}
	e^{-r^2/2}
	L_q^{(m)}(r^2) \qquad (m=p-q\ge0).
\end{equation}
The elliptical-basis versions are taken to be
\begin{equation}
\psi^E_{pq}(\boldtheta) \equiv e^{-2\mu}\psi_{pq}({\bf E}^{-1}\boldtheta).
\end{equation}
Note we have tweaked the normalizations so that the flux is
\begin{equation}
f = \int d^2\theta I(\boldtheta) = \sum_{p,q\ge0} b_{pq} \delta_{pq}
\end{equation}
rather than making the functions orthonormal.  The Gauss-Laguerre (GL)
functions are still, however, a complete and orthogonal set over the
plane.  The functions are equivalent to the ``polar shapelets'' of
\citet{Massey}. 

The advantages of the GL functions as our basis are:
\begin{itemize}
\item The $\psi_i$ are rapidly calculable via recursion relations.
\item Decompositions of typical ground-based PSFs and exponential-disk
  galaxies are reasonably compact.
\item There are obvious choices for the circularity tests: the
  centroiding condition is $b_{10}=0$; the circularity condition is
  $b_{20}=0$; and a size-matching condition is $b_{11}=0$.  The matrix
  ${\bf M}$ is very simple, {\it i.e.} it just restricts \boldb\ to five
  relevant (real-valued) elements.
\item The generator matrices ${\bf G}_k$ are sparse and simple.  In
  particular, if the truncated \boldb\ extends to $p+q\le N$, then
  ${\bf G}_{(pq)(p^\prime q^\prime)}$ vanishes for $p^\prime +
  q^\prime > N+2$.  This means that the augmented \hboldalpha\ and
  \hboldbeta\ in 
  \eqq{dbde1} can be calculated exactly simply by extending the sums
  in \eqq{sumpix} to order $N+2$ (in $p+q$), just slightly beyond the
  order $N$ needed to calculate \boldb.
\item Matrices for finite transformations are calculable with fast
  recursion relations.
\item Convolution by a PSF is also expressible as a matrix operation
  over the \boldb\ vector, if we also express the PSF as a GL
  decomposition $\boldb^\star$ in some basis ellipse ${\bf E}^\star$.
  This means that the $\phi_i$ corresponding to a chosen PSF are
  rapidly calculable.
\end{itemize}
BJ02 contains details on many of these properties, and gives the
generators (\S6.3.2), the forms for the finite-transformation matrices
(Appendix A), and the convolution matrices (Appendix B).  We
emphasize that other basis sets may be chosen which yield valid shape
measurements, but we choose the GL decomposition for its efficiency
and clarity.  Similarly, one need not execute the operation
$C(\psi_{pq})$ using the GL matrices; this is just a computational
convenience in our case.

\subsection{Alternative Circularity Tests}
We follow BJ02 in using $b_{20}=0$ as our definition of a circular
object.  It is possible that other GL coefficients carry shape
information that could be used to produce lower-noise shape estimates
\citep{Massey}.  In fact with the formalism used below
(\S\ref{errors}) to estimate shape uncertainties, one can derive for
each object a linear circularity test $M$ that is optimally sensitive to
shear.  We have implemented a scheme whereby such an optimally weighted
combination of all the quadrupole coefficients $(m=p-q=\pm2)$ is
derived for each galaxy.  Our testing has not yet demonstrated any
significant advantage to this scheme, however, so we have not pursued
it further for fear that such custom estimators might produce subtle
noise-rectification biases.

\subsection{Error Estimates}
\label{errors}
We may produce uncertainties on the ellipse parameters $E_k$ by
propagating the pixel flux uncertainties.  The covariance matrix for
the \boldb\ vector takes the usual form ${\bf C}^b = \alpha^{-1}$.  The
covariance matrix for the ellipse parameters then follows:
\begin{equation}
\label{cove}
{\bf C}^E = \left( {d\boldb \over d{\bf E}} \right)^{-1}
 {\bf C}^b \left[\left( {d\boldb \over d{\bf E}} \right)^{-1}\right]^T
\end{equation}

We may also assign a detection significance to the object.  In the
simplest case, where there is no PSF to consider, the signal-to-noise
ratio $S/N$ for
detection by the optimally-sized elliptical Gaussian filter is given
by
\begin{equation}
\nu^2 = { b_{00}^2 \over {\rm Var}(b_{00})}.
\end{equation}
The maximization of this quantity with respect to the size, shape, and
center of the elliptical-Gaussian filter generates conditions that are
identical to our default circularity tests.

We define a significance in the more general case---such as when the
PSF is non-trivial---to be
\begin{equation}
\label{nu2}
\nu^2 = { f^2 \over {\rm Var}(f)}, \qquad f = \sum_p b_{pp}.
\end{equation}
The variance on the flux can be obtained by contracting ${\bf C}^b$.
This estimate of the significance corresponds to the signal-to-noise
on the galaxy detection that would be obtained by applying a filter to
the {\em deconvolved} image that is matched to the shape, location,
and radial profile of the galaxy.  The maximum order of the sum in
(\ref{nu2}) determines the allowable complexity of the filter.

\section{Description of \glfit}

\glfit\ is the {\tt C++} implementation of the methods described in the
previous section.  The code includes classes to represent the vectors,
transformations, and covariance matrices over quantities indexed by
the integer pairs $pq$.
To fit to a single galaxy, the code requires as input:
\begin{itemize}
\item Image-format data for both the surface brightness $I_p$ and the weight
  $1/\sigma^2_p$ ({\it cf.} Eq.~(\ref{chisq})).  
  The latter image should be set to zero at saturated,
  defective, or contaminated pixels.
Data/weight image pairs from multiple exposures may
  be submitted for simultaneous fitting of a single object.  It is
  assumed that each image has been flat-fielded so that a source of
  uniform surface brightness has equal value in all pixels.
\item For each exposure, a map $\boldtheta(\boldx)$ from
  the pixel coordinate 
  system into the world coordinate system.  This is typically either
  the distortion map into the local tangent plane to the celestial
  sphere, or an identity map.
\item For each exposure, the local sky brightness 
  and a photometric gain factor. 
\item If we
  are executing a {\em deconvolution fit}, whereby we measure 
  the shape of the galaxy before PSF convolution, then we require,
  for each
  exposure, a vector $\boldb^\star$ and basis ellipse ${\bf E}^\star$
  describing the estimated PSF at the location of the galaxy.
  If we are executing a {\em native fit}, whereby we are trying to
  measure the {\em observed} shape of the galaxy, then no PSF
  information is required.
\item A choice for the initial order $N$ $(p+q\le N)$ of the GL
  decomposition of the galaxy model.
\item Lastly an initial estimate of the size, location, and
  shape of the galaxy, {\it e.g.\@} the values generated by detection
  software such as {\tt SExtractor} \citep{Bertin}.  If we are
  executing a deconvolution fit, we need to know both the size
  $\sigma_o=e^{\mu_o}$ of the object as observed, and the size
  $\sigma_\star$ of the PSF in which the object is observed.  These
  are by default subtracted in quadrature to estimate the size
  $\sigma_i$ of the intrinsic (pre-seeing) galaxy.
\end{itemize}

The procedure for shape determination has the following steps:
\begin{enumerate}
\item If we are executing a deconvolution fit, we choose the size of
  the ``source basis'' ${\bf E}_s$ which is
  the basis ellipse for the $\psi^E_i$ that will be used to model the
  pre-convolution image.  We select
\begin{equation}
e^{2\mu_s} = \sigma^2_s = \sigma^2_i + f_{\rm PSF} \sigma^2_\star,
\end{equation}
where $\sigma^2_i$ and $ \sigma^2_\star$ are the sizes of the
pre-seeing galaxy image and the PSF, respectively, and $f_{\rm PSF}$
is a prefactor chosen by default to be 1.2.  If there are multiple
exposures, we take 
$\sigma_\star$ to be the harmonic mean of all the exposures' PSF sizes.
As discussed in BJ02 \S6.2, we expect a choice
of Gaussian basis $\sigma$  that is larger than both the intrinsic
galaxy and the PSF to offer the most sensitive measure of
pre-convolution quadrupole moments.  During a deconvolution fit,
$\mu_s$ is held fixed during all iterations and there is no constraint
on size in the circularity test ${\bf M}$.

\item The initial guess for ${\bf E}_s$ is mapped from the world
  coordinate system into each exposure's pixel system.  An elliptical
  mask is created for each exposure, defined by the $n\sigma$ contour
  of the elliptical Gaussian.  Typically we select $n\approx4$.  Pixels
  outside the mask are not included in the fit.  Larger masks reduce
  the possibility of degeneracy or non-positive models for high-order
  fits, but increase the 
  chance of contamination by neighboring objects.

\item If any of the exposures do not fully contain the fitting mask, 
  the {\tt Edge} flag is set.  If the initial centroid
  falls outside {\em all} the exposures, then the {\tt OutOfBounds}
  flag is set and the fit is terminated.

\item The basis ellipse ${\bf E}^\star$ of the PSF is changed to have
  the same shape (not size) as the current estimate of the galaxy
  shape, by applying a finite shear transformation matrix to
  $\boldb^\star$.  The convolution by the PSF is now expressed as a
  matrix operation, so we have an observed-plane image model
\begin{equation}
  I_o(\boldx) = \sum_{i,j} \psi^{E_o}_i[\boldtheta(\boldx)] C_{ij} b_j .
\end{equation}
  The observed-plane basis ellipse ${\bf E}_o$ has second moments that
  are the sum of the second moments 
  of the chosen source-plane basis ${\bf E}_s$ and the PSF basis
  ellipse.  The convolution matrix elements $C_{ij}$ can be computed
  as in Appendix B of BJ02.  We now have a model for the observed data
  which is linear over the expansion coefficients \boldb.
\label{psfstep}

\item The linear solution for \boldb\ is executed by summing over all
  valid pixels within the masks on all exposures.  If a singular value
  decomposition of the \boldalpha\ matrix indicates that there are
  poorly constrained combinations of \boldb\ elements, the order of
  the GL decomposition is reduced by one (to a minimum of $N=2$), the
  {\tt ReducedOrder} flag   is set, and the solution is attempted
  again.

\item The matrix $d\boldb / d{\bf E}$ is calculated from the augmented
  \hboldalpha\ and \hboldbeta\ (which in fact were calculated in the
  previous step) as per \eqq{dbde1}.

\item If the circularity test applied to \boldb\ is {\em not} zero within a
  specified tolerance, then an iteration increment $\delta {\bf E}$ is
  calculated according to \eqq{deltae}.  If a
  singular-value decomposition of $d\boldb / d{\bf E}$ indicates a
  nearly-degenerate matrix, we know that there is no sensible estimate
  for the next step.  In other words, at least one of our circularity
  tests is currently ill-defined, for example no small change of basis
  makes the object appear more circular. This should, for example, be
  the case for a point source, but also arises for some very irregular
  or noisy galaxies.  
  The behavior of the next step depends upon the fit type:
  In a native fit,
  we set the {\tt FrozeDilation} flag, set $\mu$ back to its
  initial value, and restart the fitting process, this time holding
  $\mu$ fixed.  If we are executing a
  deconvolution fit, we increment $f_{\rm PSF}$ by 0.5, set the 
  {\tt RaisedMu} flag, and restart the
  fitting process, since fits to a larger basis tend to be more stable
  (but less sensitive).  If 
  {\tt FrozeDilation} has been set (for a native fit) or
  $f_{\rm PSF}>3$ already (for a deconvolution fit), 
  and $d\boldb / d{\bf E}$ is still near-degenerate,
  then our last resort is to set the {\tt 
  FrozeShear} flag, reset \boldeta\ to its starting value, and
  restart the fit process while holding the shape fixed. 

\item The next iteration of the fitting process continues at
  step~\ref{psfstep}.  If the circularity test is not satisfied within a
  chosen number of iterations, the {\tt DidNotConverge} flag is set
  and the fit is terminated.  If any other matrix singularities are
  encountered, or if there are too few pixels to conduct the fit, the
  {\tt Singularity} flag is set, and the fit terminates.

\item  Once the circularity tests are satisfied within a desired
  tolerance, the resultant ${\bf E}$ and \boldb\ for the galaxy are
  available for the user.  The \boldalpha\ matrix provides the
  covariance matrix for \boldb.  The errors on the centroid, size, and
  ellipticity (or whichever of these were free to vary) are calculated
  using \eqq{cove}, and a detection significance can be reported using
  \eqq{nu2}. 
\end{enumerate}
In most cases the solution for ${\bf E}$ is executed twice: first, a
coarse fit at order $N=2$ is attempted.  If this fails, the {\tt
  CoarseFailure} flag is set.  If it succeeds, the coarse solution is
used as the starting point for a fit to the full requested $N$.

Another subtlety is that the pixel mask is not changed between
iterations---the mask maintains the shape and size specified by the
initial guess (or the result of the coarse fit).  This should not bias
the fit toward the initial guess, however, because we are fitting to
the data within the mask rather than executing a sum of moments over
pixels within the mask.  If the size $\mu$ of the object changes
dramatically during the coarse fit ({\it e.g.} due to a very poor
initial estimate from {\tt SExtractor}), then we do resize the masks
and start over.  If at any time the mask on an image grows to
encompass too large a number of pixels, we set the {\tt TooLarge} flag
and quit.  This prevents the program from getting hung up calculating
pixel sums for nonsensically large objects.

The algorithm is considered to have converged to a valid shape
measurement if it completes without setting of the flags
{\tt FrozeShear, DidNotConverge, Singularity, OutOfBounds,
TooLarge,} or {\tt CoarseFailure}.

\section{Test Procedure}

In our tests, postage-stamp images are created of individual galaxies,
as convolved with a known PSF, pixelized, and given noise.  We then
extract shape and shear measurements from ensembles of such
postage-stamp tests. In the real sky and in end-to-end tests which
draw (pre-lensing) galaxy shapes at random, the uncertainties in the
output shear are 
usually dominated by the finite variance of the mean pre-shearing
shape (shape noise). We construct our ensembles with zero mean shape
to eliminate shape noise, leaving our testing precision dominated by
the shot noise in the images.

Because we provide the shape-measurement algorithms with an
approximate location for every simulated galaxy, 
our tests do not include the effects of possible selection
biases \citep{K00,BJ02,Hirata03}.  Similarly the PSF is always
presumed to be known exactly, so we do not test for errors that would
result from PSF measurement or interpolation errors \citep{HoekstraXX,
Jarvis06}.  Effects of galaxy crowding or overlap, detector
nonlinearities, intrinsic galaxy-shape correlations, or redshift
determination errors \citep{Ma} are not examined
here.  Instead we concentrate on the sources of systematic error that
arise specifically from (1) the shape measurement,
(2) removal of the PSF effects, and (3) the shear estimation,
in images with finite noise and sampling.

\subsection{Test Conditions}

In order to test our WL analysis methodology, 
we need to find the conditions under which \glfit\ works,
and then test for any biasing in the shape measurement or
shear estimation method.
The following is the list of the test conditions or choices made:

\begin{itemize}
\item Native fit or deconvolution fit.  
All measurements are tested for cases without (native fit) and 
with (deconvolution fit) the presence of a PSF.
In a WL analysis of real sky data, native fits are used in
(1) characterizing the PSF from stellar images, and
(2) measuring the shape of the PSF-smeared galaxy, 
whose value is used to produce an initial estimate of the intrinsic
galaxy shape.
\item Galaxy size with respect to pixel or PSF.
As the pixel sampling rate decreases, it becomes harder to obtain 
shape information about the galaxy from the pixelized image,
or for the \glfit\ to converge.
For the native fits, the galaxy size is defined with respect to the pixel;
for the deconvolution fits, to the PSF.
\item Galaxy detection significance $\nu$, or equivalently $S/N$.
If the background noise dominates the signal from the galaxy,
\glfit\ is also unlikely to converge.
\item Galaxy shape $e$.
The galaxies are sheared to various shapes, ranging from
$|e|=0$ to $0.9$ ($=$ axis ratio $4.35:1$).  
In some cases, the major axis orientation
(parallel or diagonal to the pixel axis) is also examined.
\item Galaxy type.  We choose a ``symmetric'' or ``asymmetric''
galaxy model, with the latter being a more stringent test of the
methodology. 
The galaxy models are described further in 
\S\ref{imagemodel}, and the shear accuracy is examined by the ring test,
described in \S\ref{ringtest}.
\end{itemize}
For the deconvolution fit, we have the following additional choices 
for the PSF:
\begin{itemize}
\item PSF type.  We choose an Airy disk PSF as the worst possible 
case for a PSF.  An Airy disk offers particular difficulties for
shape-measurement algorithms, such as 
its non-trivial morphology and divergent second moments.  
\item PSF shape.  The Airy disk is either circular or
elliptical (sheared at $e_{\rm PSF}=0.1$).  We look for any leakage of the 
PSF shape into the galaxy shape measurement or the shear estimate.
\end{itemize}

In this paper, the PSF is always known; 
we do not test for errors due to incomplete knowledge of the PSF.

\subsection{Pixelized Galaxy Models}
\label{imagemodel}

Two types of galaxies, symmetric and asymmetric, are used
to test the shape measurement and shear recovery.  The galaxy models,
along with any shear, rotation, and dilation, are described 
as an analytic function over the pixel coordinates. 

The ``symmetric'' galaxies have either a Gaussian or exponential 
circularly symmetric radial profile that is then sheared.  All
isophotes are similar ellipses, so these galaxies have
unambiguous shapes, which allows direct comparison between the
input and measured shape.
The elliptical symmetry of these galaxies could be masking errors in
the shape-measurement methodology by causing fortuitous cancellations
of errors.  In other words, all of their EGL expansion coefficients
with $m\ne 0$ are exactly zero, so any biases that couple to these
coefficients will not be tested by the symmetric galaxies.

The ``asymmetric'' galaxies are designed so that they have no 
inversion symmetry, and their isophotal shape varies with radius; i.e.,
there is no unambiguously defined shape.  The ring test is therefore
needed to see if input shear is accurately recovered.
These are constructed as a combination of exponential and deVaucouleur
ellipses, each with different flux, size, and shape, as well as
a slight offset between the two centroids.  
The asymmetric galaxies offer a more stringent test of shear
measurement, since we have broken any symmetries that could mask
errors in the shape measurement.

A pixelized postage stamp image is created from the analytic functions 
that describe the galaxy and PSF models at the intended sampling rate.
The centroid of the analytic functions is randomized within 
the pixel over multiple image realizations (pixel phase randomization).
The PSF convolution is done either analytically if possible,
or via FFT of noiseless pixelized images.
The Poisson photon noise is then added to produce the final image
to produce the desired $S/N$ for the object.
We then run \glfit\ for shape measurement with starting 
parameters  randomly displaced from the true values.

\subsection{The Ring Test}
\label{ringtest}

The ring test evaluates the mean shape---the weighted average 
of the galaxy shape components---with a simplified 
galaxy shape distribution, as seen in Figure~\ref{fig:ringtest}. 
When a shear \bolddelta\ is uniformly applied to the ensemble, 
the mean shape $\langle \bolde \rangle$ should equal ${\mathcal
  R}\bolddelta$, where 
${\mathcal R} = 1-\frac{e^2}{2}$ (Eq.~(\ref{avgweakmap})) is 
called the responsivity.
This relation then gives an estimate of the shear from the mean shape.

The value of the ring test is two-fold: 
the expected signal is known, and the shape noise is absent.
The shape noise in WL is the statistical noise due to the random 
distribution of shapes with finite magnitudes; 
the uniform distribution of shapes in the ring test eliminates this noise.  
The lack of shape noise and the known expected value then
allow systematic errors to be quantified.

\section{Results}

\subsection{\glfit\ Convergence}

We run \glfit\ on objects with different size and 
detection significance to understand the conditions under which 
\glfit\ converges reliably, which we define to be $>99\%$ 
rate of successful shape measurement.
Shear estimate accuracy tests are subsequently
performed only in $>99\%$ convergence conditions, to preclude any 
selection biases that could be generated by 
the non-convergence of \glfit.

\subsubsection{Native Fit Convergence}
\label{nativeconvergence}

\begin{figure}[!tbp]
\plotone{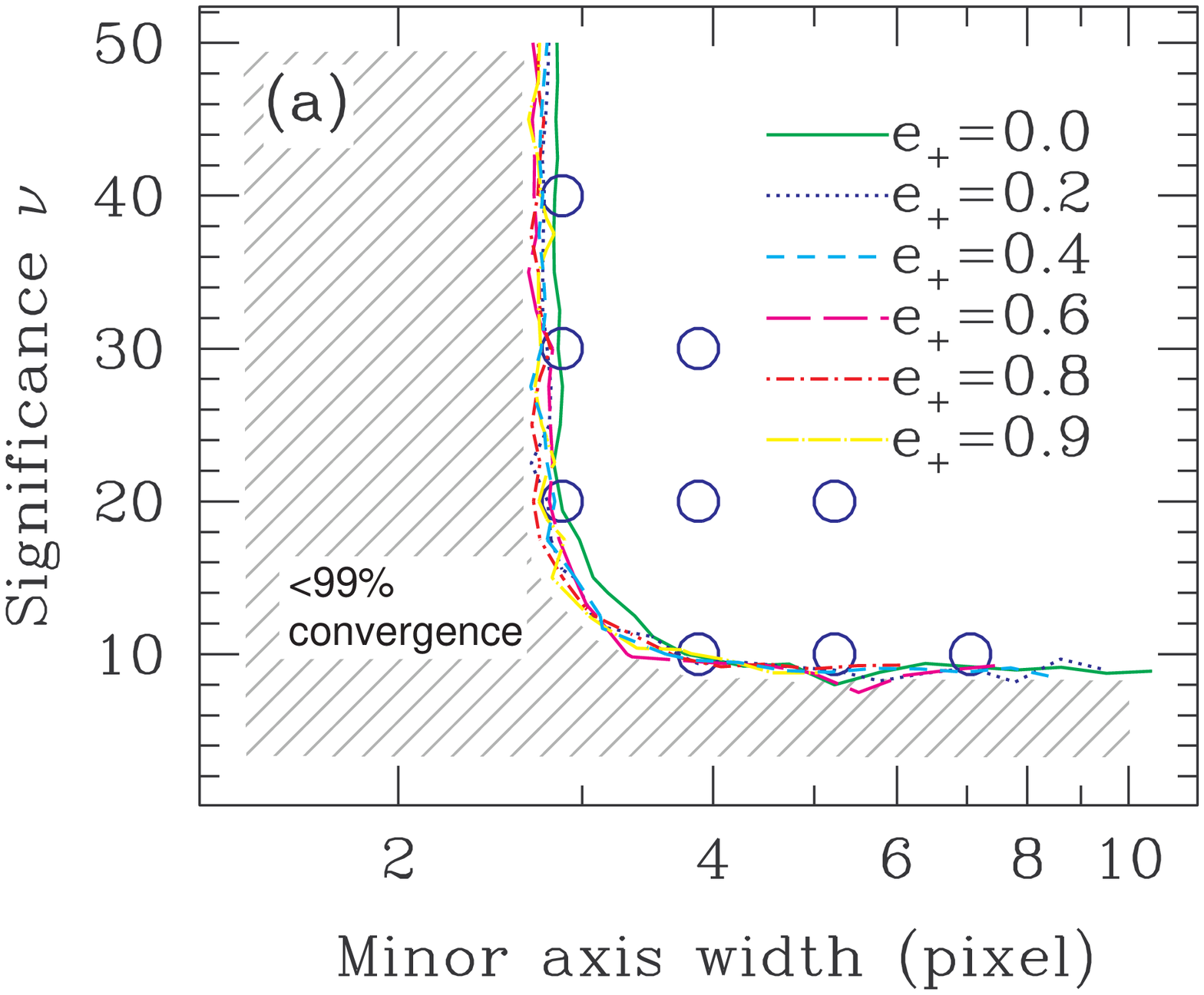}
\plotone{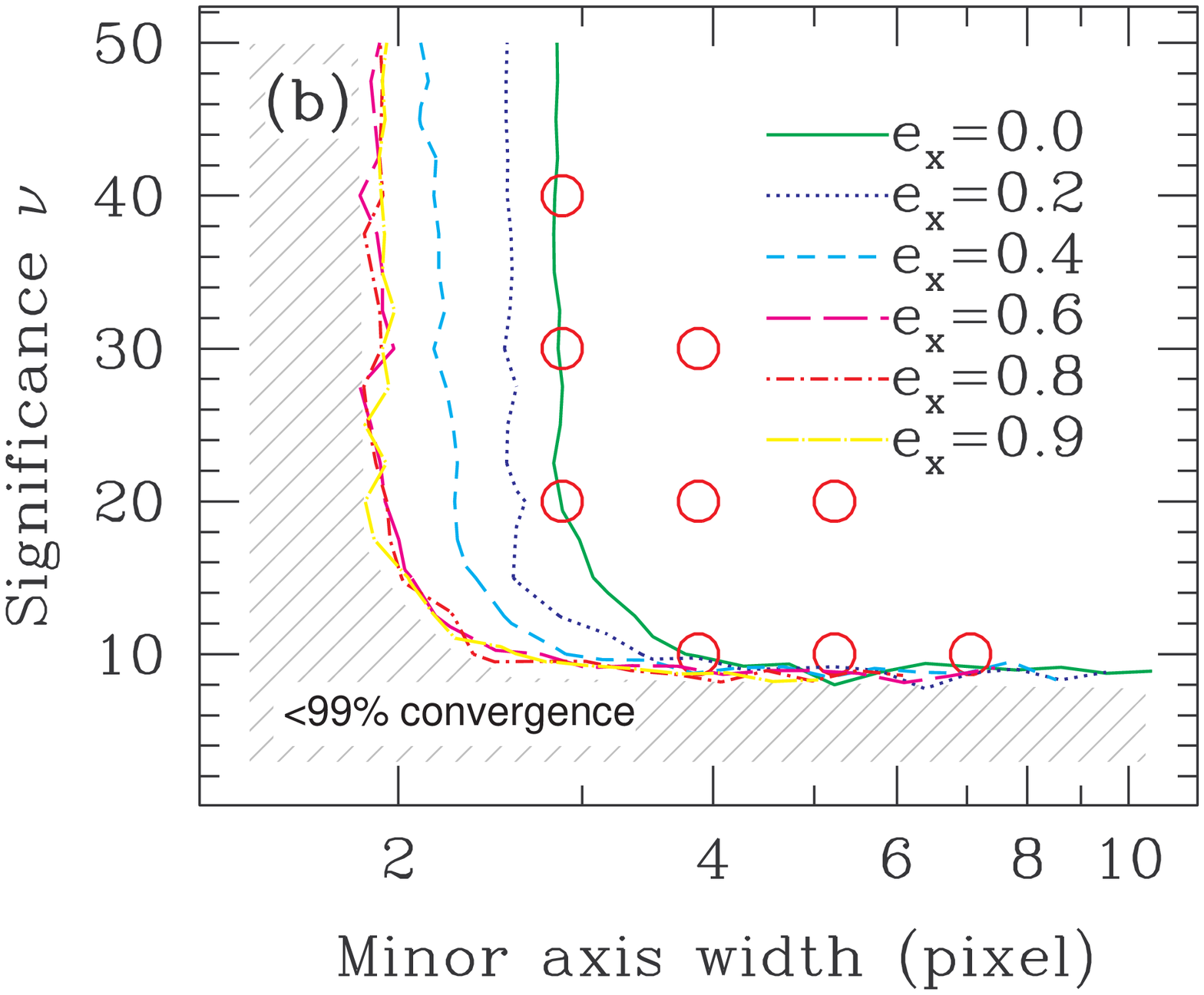}
\figcaption{Contours of 99\% successful \glfit\ native-fit convergence
over pixel sampling (horizontal axis) and object $S/N$ (vertical axis).
The pixel sampling is defined as the FWHM along the minor axis direction,
or the ``minor axis width''.
The shaded region is where the native-fit convergence rate is $<99\%$.
(a) Convergence when major axis is along the pixel axis.
(b) Convergence when major axis is along the pixel diagonal.  
The shift in the contour in (b) is due to an effectively new pixel spacing
({\em cf.} Fig.~\ref{fig:diag-sampling}).
The circles in each plot indicate the parameters at which 
the shape measurement accuracy tests have been performed 
({\em cf.} Fig.~\ref{fig:nativeshapeerror}).
\label{fig:native-contour}}
\end{figure}

Figure~\ref{fig:native-contour} shows the $99\%$ convergence contour 
for native fits on exponential-profile symmetric galaxies of various
ellipticities.
The upper right hand region delineated by the
contour indicates where the fit has $>99\%$ convergence.
The horizontal axis indicates the sampling rate, and is expressed 
in terms of the ``minor axis width'' (MAW), defined below.
The vertical axis incidates the detection significance $\nu$ of the object.

We define the minor axis width for a Gaussian object as its 
full width at half maximum (FWHM) along the minor axis direction.
For non-Gaussian objects, the ``Gaussian size'' $\sigma$ is defined 
from the size-matching condition $b_{1\!1}=0$ (\S\ref{gldecomposition});  
an exponential profile $I(r) = I_0 \exp(-r/r_0)$ has a Gaussian
size $\sigma \simeq 1.16r_0$.
The Gaussian size along the minor axis is
$\sigma_b=\sigma\exp(-\eta/2)$, where $\eta$ is the conformal shear
(\S\ref{shape-as-shear}), so the minor axis width is 
$2.35\sigma_b$.

Figure~\ref{fig:native-contour}(a) clearly shows that 
for a native fit to converge reliably, $S/N$ must be $\geq10$ and 
its minor axis width be $\geq2.8$ pixels.
The convergence contour for Gaussian radial profiles is found to
be identical to the exponential-profile contour.
The fact that the minor axis dimension is the limiting factor for 
convergence implies a shear-dependent selection, since 
$\sigma_b$ does not remain constant under a shear operation.

\begin{figure}[!tbp]
\plotone{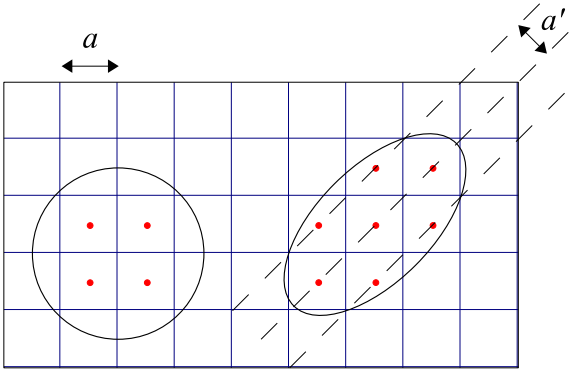}
\figcaption{Relation of the minimum measureable object size
(Fig.~\ref{fig:native-contour}) to the pixel lattice for native \glfit.
The circle and the ellipse have the minimum minor-axis width
for reliable shape measurement in native fits.
For the ellipse on the right ($e_\times=0.6$), the sampling spacing
along the minor axis, $a'$, is effectively $1/\sqrt{2}$ of the pixel
lattice spacing $a$.
The sampling spacing is still $a$ for ellipses stretched along the pixel axis.
We note that the the object center is placed at random locations within
the pixel in our simulations.
\label{fig:diag-sampling}}
\end{figure}

The required sampling rate depends, however, on the 
orientation of the minor axis.
Figure~\ref{fig:native-contour}(a) shows the convergence contour 
when the ellipse is elongated along the pixel axis,
where Figure~\ref{fig:native-contour}(b) shows the case 
when the elongation is along the pixel diagonal.
In the second case, the required sampling rate appears to decrease with 
increasing $e$, until it settles at 2.0 pixel for $e\geq0.6$.  
Figure~\ref{fig:diag-sampling} illustrates
that for objects elongated along the pixel diagonal,
the effective sampling rate along the minor axis becomes $1/\sqrt{2}$
of the pixel spacing.

The GL decomposition of the PSF is determined via a native fit.
Hence, the images must sample the PSF minor axis width by
more than 2.8 pixels for the \glfit\ to be applied successfully
to individual stars.
Note that it is possible to increase the sampling rate by dithering
for a CCD of any pixel size.

\begin{figure}[!tb]
\plotone{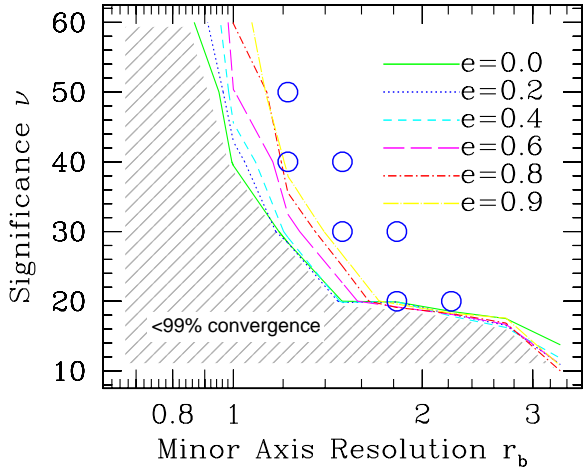}
\figcaption{Contour for 99\% deconvolution-fit shape measurement 
convergence over resolution (horizontal axis) and object $S/N$ 
(vertical axis), for an Airy PSF convolution.
The shaded region is where the deconvolution-fit convergence rate is $<99\%$.
The resolution is expressed as $r_b\equiv\sigma_b/\sigma_{\rm PSF}$, where
$\sigma_{\rm PSF}$ and $\sigma_b$ are the characteristic Gaussian widths
for the PSF and along the galaxy minor axis, respectively.
The contour is independent of the orientation of the major axis.
The circles indicate the parameters at which 
the shape measurement accuracy tests have been performed 
({\em cf.} Fig.~\ref{fig:dcvlshapeerror}).
\label{fig:deconv-contour}}
\end{figure}

\subsubsection{Deconvolution Fit Convergence}
\label{deconvconvergence}

For the deconvolution test, we convolve the model galaxies with
a circular Airy-function PSF, whose characteristic
width is $\sigma_{\rm PSF}\simeq 0.21\,\lambda/D$.
The model galaxies in this test are symmetric elliptical objects of 
exponential radial profile with minor axis Gaussian width $\sigma_b$ 
before PSF convolution.  The pixel size is irrelevant to 
the convergence rate, as long as $\sigma_{\rm PSF} \ge\,2.8$~pixels
({\em cf.} \S\ref{nativeconvergence}).
We find that the convergence rate of \glfit\ is independent of the
orientation of the major axis under these conditions.

Figure~\ref{fig:deconv-contour} shows the 99\% convergence contour
for deconvolution fits.  The horizontal axis is 
the minor-axis resolution 
\begin{equation}
r_b\equiv\sigma_b/\sigma_{\rm PSF}.
\end{equation}
We find that high-$e$ galaxies are less
likely to converge at low resolution than more circular
galaxies.  

In analyses of the real sky, the dimmest and smallest galaxies are
most numerous, so we draw our attention to
the lower left corner of the contour in Figure~\ref{fig:deconv-contour}.
\glfit\ will 
converge reliably for objects with
$r_b\geq\,1.8$ down to $\nu\ge 20$.
To obtain a reliable shape measurement of marginally resolved objects
($r_b\geq 1.2$) at all shapes $e$, the significance must be
$\nu\ge40$.

\subsection{Accuracy for Symmetric Galaxies}

Having determined where \glfit\ is consistently successful
in measuring a shape, we now
examine the accuracy of the shape measurement 
and error estimate that result.

When an elliptical object is used as an input, the 
input shape is well defined, and hence so is the measurement error.
In this section we use symmetric elliptical galaxies with 
exponential radial profile that have an unambiguous shape $\eta_{\rm input}$.
The shape measurement error is expressed as a conformal shear
$\Delta\eta\equiv(-\eta_{\rm input})\oplus\eta_{\rm meas}$, where
$\oplus$ is the shear addition operator, equivalent to shear matrix 
multiplication:
\begin{equation}  
\boldeta_3 = \boldeta_2 \oplus \boldeta_1 
\ \ \ \ \Leftrightarrow \ \ \ \ 
{\bf S}_{\boldeta_3}{\bf R} = {\bf S}_{\boldeta_2}{\bf S}_{\boldeta_1},
\end{equation}  
where ${\bf R}$ is the unique rotation matrix that allows
${\bf S}_{\boldeta_3}$ to be symmetric (BJ02 \S2.2).
The quantity $\Delta\eta$ is the amount of shear required 
to bring the input shape to the measured shape, hence describes a
shear bias regardless of the intrinsic object shape.
Each data point is an average over $N=100,000$ measurements of 
independent realizations of pixel phase and photon noise.
We then plot the averaged error components 
$\langle\Delta\eta_+\rangle$ or $\langle\Delta\eta_\times\rangle$
as the measurement accuracy.
The uncertainty in the mean $+$ component 
is then $\sigma_{\langle\eta_+\rangle}=\sqrt{\sum(\Delta\eta_+)^2/N^2}$.

\glfit\ produces error estimates $\tilde{\sigma}_{\eta_+}$, 
$\tilde{\sigma}_{\eta_\times}$ for every shape measurement 
from the covariance matrix ${\bf C}^{E}$ (\S\ref{errors}).
We test for the accuracy of this estimation by examining the averaged ratios
$\langle(\Delta\eta_+)^2/\tilde{\sigma}^2_{\eta_+}\rangle$ and
$\langle(\Delta\eta_\times)^2/\tilde{\sigma}^2_{\eta_\times}\rangle$, where
the average is over shape measurement trials.

\begin{figure}[!tb]
\plotone{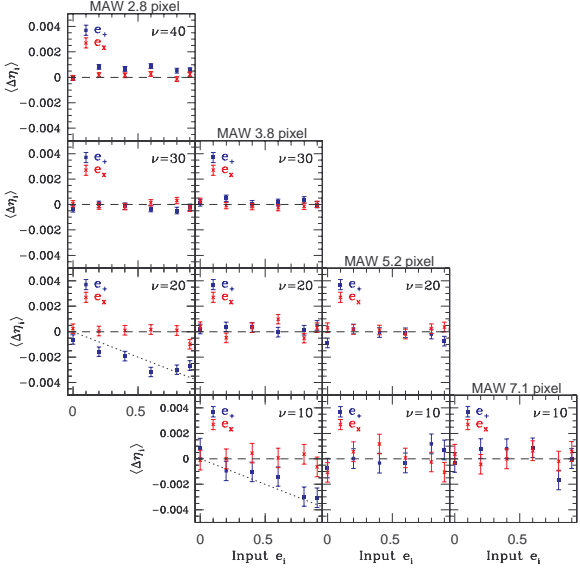}
\figcaption{The error in native fit shape measurement as a function 
of the input shape $e$.  The ``shape error'' $\Delta\eta$ is the
amount of shear necessary to bring the measured shape to the 
input shape; $\Delta\eta_i$ is the $i$th component of the error.
The plots in the same column have the same minor axis width 
(MAW$\equiv2.35\sigma_b$), 
whereas the plots in the same row have the same significance $\nu$;
each panel correspond to the set of $(\nu,{\rm MAW})$ parameters
indicated by the circles in Figs.~\ref{fig:native-contour}(a) and (b).
The slanted, dotted line in two of the panels indicate
shape error of $\Delta\eta_i/e_i=-0.4\%$, where $i=+,\times$.
\label{fig:nativeshapeerror}}
\end{figure} 

\begin{figure}[!tbp]
\plotone{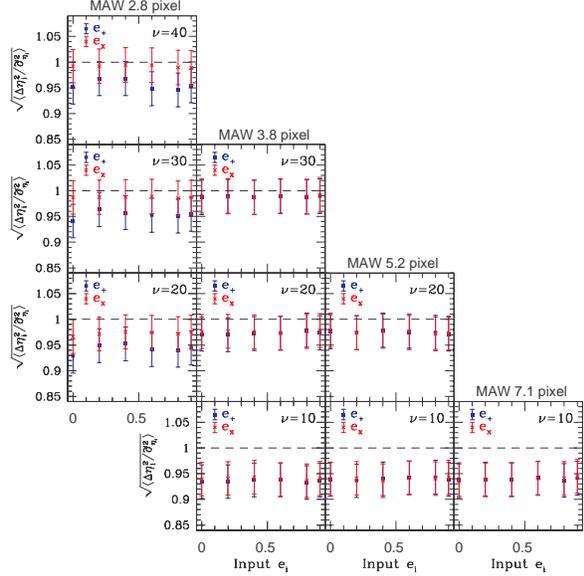}
\figcaption{The RMS average of the ratio of measured error to estimated error.
The error estimate $\tilde{\sigma}_{\eta_i}$ is obtained from 
the fitting process ({\em cf.} \S\ref{errors}).
Each panel correspond to the set of parameters indicated by the circles in
Figs.~\ref{fig:native-contour}(a) and (b). 
\label{fig:nativeerrorestimate}}
\end{figure}

\subsubsection{Native Fit Errors}
\label{nativeerror}

Each panel in Figure~\ref{fig:nativeshapeerror} show the average error
$\langle\Delta\eta_i\rangle$ ($i=+,\times$) as a function of the input 
shape $e$ of the symmetric exponential galaxies.  We first examine the 
accuracy of native fits to galaxies with no PSF convolution.
The panels correspond the sampling rates and detection significances
marked by the points inside the 99\% convergence contour in
Figure~\ref{fig:native-contour}.

Those that are well into the $>99\%$ convergence region 
(the inner three panels) show no systematic bias in the shape measurement
within the uncertainty $\sigma_{\langle\eta_i\rangle} \simleq\, 0.0005$.
For those on the $99\%$ convergence contour (left-most and bottom panels),
shape measurements show some systematic errors, but the worst systematic is
still within $\Delta\eta_i/e_i\simeq0.4\%$.
The shear error $\Delta\gamma/\gamma$ ($=m-1$ if $c=0$) 
that would result from such measurement error is
\begin{equation}
\frac{\Delta\gamma}{\gamma} \ \simeq \ f \frac{\Delta\eta_i}{e_i},
\end{equation}
assuming $\Delta\delta/\delta\simeq\Delta\gamma/\gamma$, where 
\begin{equation}
f \simeq \left[\frac{1+\langle e^2\rangle/2}{1-\langle e^2\rangle/2}\right]
(1-\langle e^2\rangle)
\end{equation}
for a constant $\Delta\eta_i/e_i$.
The function $f$ is always $<1$, and rapidly decreases to zero for 
$0.6\,\simleq\,e<1$.
Hence a shape measurement systematic of 
$\Delta\eta_i/e_i\simeq0.4\%$ approximately corresponds to 
a shear calibration error of $\Delta\gamma/\gamma\simeq 0.4\%$.

Figure~\ref{fig:nativeerrorestimate} shows the root-mean-square (RMS)
of the actual to estimated error ratios,
$\sqrt{\langle\Delta\eta_i^2/\tilde{\sigma}_{\eta_i}^2\rangle}$, 
where $\tilde{\sigma}_{\eta_i}$ is the error estimate in the $i$th 
component as calculated by \glfit.  Overall, we see that the error 
estimate does fairly well.  There is a slight tendency to overestimate 
the error as $\nu$ becomes low;  
the $\tilde{\sigma}_{\eta_i}$ bias has no dependence on the pixel sampling.
The $\nu$ dependence is approximately 
$\sqrt{\langle\Delta\eta_i^2/\tilde{\sigma}_{\eta_i}^2\rangle}\simeq $
$1-\frac{0.6}{\nu}$.

\begin{figure}[!tbp]
\plotone{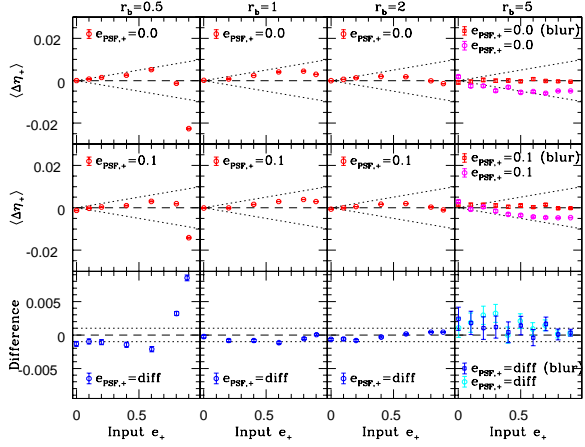}
\figcaption{The error in deconvolution shape measurement at $\nu=100$,
at resolutions $r_b=0.5$, $1$, $2$, and $5$.
The plots in the same column have the same resolution.
The top row is convolved with a circular Airy PSF; middle row with
an elliptical Airy PSF ($e_{\rm PSF}=0.1$); the bottom row is 
the difference between the two.
In the top two rows, the slanted, dotted lines indicate
shape error of $\Delta\eta_+/e_+=\pm1\%$,
where in the bottom row, the dotted
lines indicate $\langle\Delta\eta_+\rangle=\pm0.001$, or 
$1\%$ of the PSF anisotropy.
For $r_b=5$, the systematic 
specifically due to the GL decomposition of the Airy PSF 
is removed by intentionally blurring both the PSF and 
the observed image with a Gaussian that is
a fraction of the size of the galaxy.
\label{fig:dcvlshapeerror100}}
\end{figure} 

\begin{figure}[!tbp]
\plotone{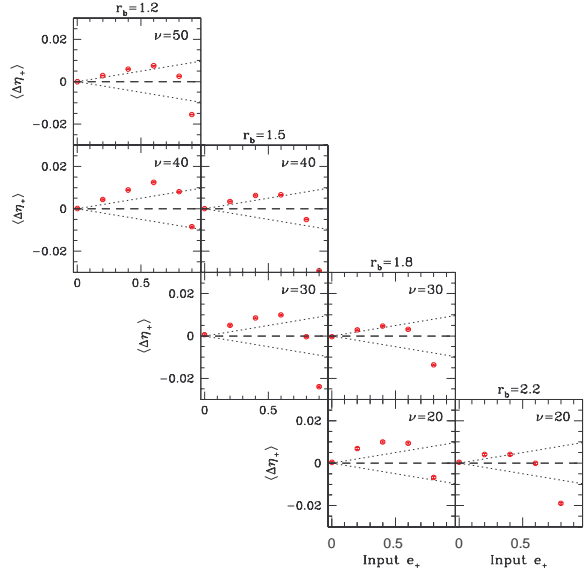}
\figcaption{The error in deconvolution shape measurement for $\nu\leq50$.
The Airy PSFs are isotropic ($e_{\rm PSF}=0$). 
The plots in the same column have the same resolution $r_b$, 
whereas the plots in the same row have the same significance $\nu$;
each panel correspond to the set of $(\nu,r_b)$
parameters indicated by the circles in Fig.~\ref{fig:deconv-contour}.
These plots are similar to those on the top row of 
Fig.~\ref{fig:dcvlshapeerror100}.
The slanted, dotted lines indicate shape error of $\Delta\eta_+/e_+=\pm1\%$.
\label{fig:dcvlshapeerror}}
\end{figure} 

\begin{figure}[!tbp]
\plotone{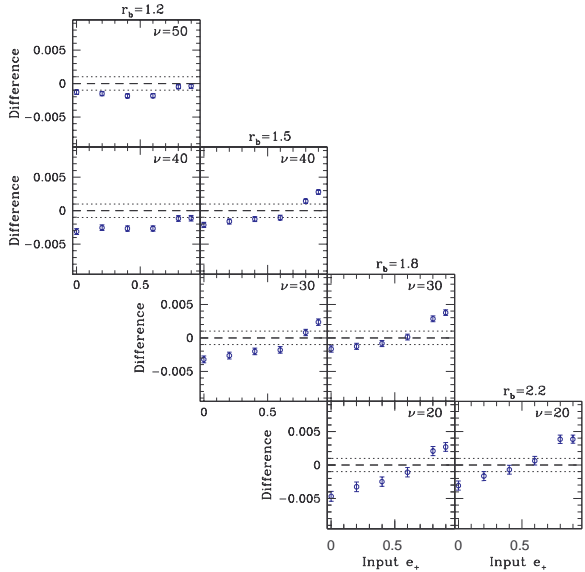}
\figcaption{The PSF anisotropy bias in deconvolution shape measurement 
for $\nu\leq50$.
Each panel correspond to the set of $(\nu,r_b)$
parameters indicated by the circles in Fig.~\ref{fig:deconv-contour}.
These plots are similar to those on the bottom row of 
Fig.~\ref{fig:dcvlshapeerror100}.
The dotted lines indicate $\langle\Delta\eta_+\rangle=\pm0.001$, or 
$1\%$ of the PSF anisotropy.
\label{fig:dcvlshapebias}}
\end{figure} 

\begin{figure}[!tbp]
\plotone{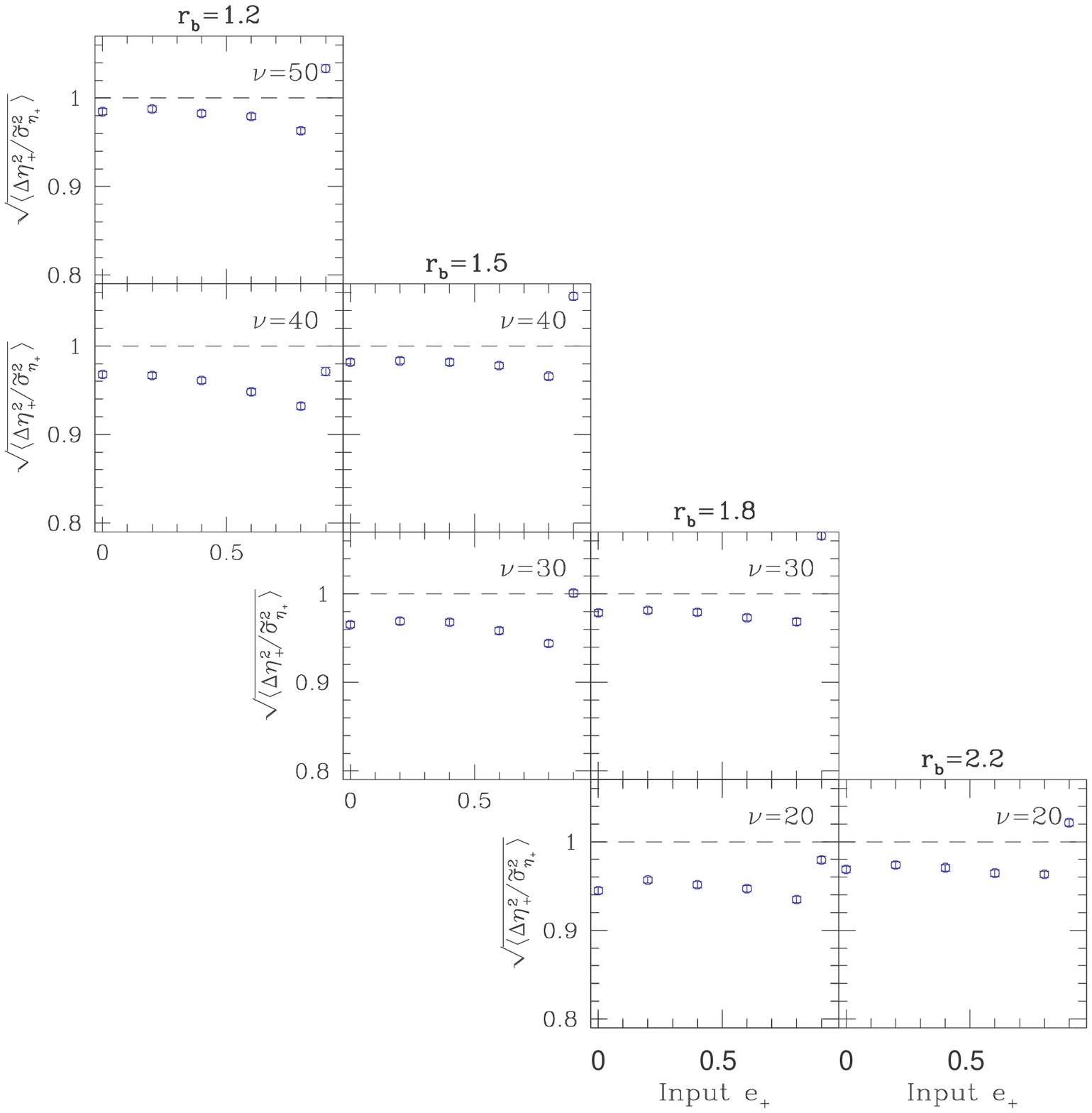}
\figcaption{The RMS ratio of actual ($\Delta\eta_+$) to estimated
($\tilde{\sigma}_{\eta_+}$) errors for deconvolution fits.
Each panel correspond to the set of $(\nu,r_b)$
parameters indicated by the circles in Fig.~\ref{fig:deconv-contour}.
\label{fig:dcvlerrorestimate}}
\end{figure}

\subsubsection{Deconvolution Fit Errors}
\label{deconverror}

Figure~\ref{fig:dcvlshapeerror100} shows the accuracy of a
deconvolution fit when a symmetric, exponential galaxy is convolved 
with an Airy PSF.  We first discuss a high-$S/N$ case, $\nu=100$.
The measurements were done at various minor axis 
resolutions, where $r_b=0.5$, 1, 2, and 5 each correspond to 
the columns from left to right.

The panels in the top two rows show the shape error 
$\langle\Delta\eta_i\rangle$ as a function of the input galaxy ellipticity.
In the top row, the Airy PSF is circular
($e_{\rm PSF}=0.0$), while in the middle row
the Airy PSF is anisotropic, with an ellipticity of $e_{\rm PSF}=0.1$.
The PSF and galaxy ellipticity both are elongated along $e_+$;
we find that the results do not change even if we add a non-zero
$e_{\times}<0.15$ component to either the PSF or the galaxy.
The major axis orientation with respect to the pixel axis 
is irrelevant for deconvolution fits ({\em cf.} \S\ref{deconvconvergence}).

The bottom row plots the difference in average measured shape
between $e_{\rm PSF}=0.1$ and $e_{\rm PSF}=0.0$.
If the PSF anisotropy is fully suppressed in the deconvolved
galaxy-shape measurement,  
this difference should be zero.

In general, Figure~\ref{fig:dcvlshapeerror100} shows the following
two systematic shape measurement errors: 
(1) Regardless of the PSF anisotropy, the measurement error 
exhibit a slight systematic when the resolution is low 
($r_b<2$).  There is a constant positive slope 
in $\langle\Delta\eta_+\rangle$ between $0<e<0.6$.
(2) For $e>0.6$, the slope turns around, eventually underestimating 
the shear at $e=0.9$. 
In the region described by (1), the shear measurement errors are
approximately $\langle\Delta\eta_i\rangle\,\simleq\,0.01e_i$,
corresponding to $\Delta\gamma/\gamma\,\simleq\,1\%$.
The well-resolved galaxies ($r_b\ge5$) exhibit biases in
shape measurement, which are a consequence of
a peculiarity of the Airy function PSF.  We will explain this below,
and describe a remedy which yields the improved performance of the points
marked ``blur.''

Despite these measurement systematics, however, 
the difference between the $e_{\rm PSF}=0.1$ and $0.0$ 
measurements are less than a shear of 0.001, equivalent to $<1\%$ of 
the PSF anisotropy, except for the poorly resolved ($r_b=0.5$) and highly
elliptical galaxies.
Hence our \glfit\ implementation has a PSF anisotropy bias suppression of 
$>99\%$ at $\nu=100$, except for the poorly-resolved case.

We now examine deconvolution fits with galaxies of lower $S/N$, 
using the combinations of resolution and significance marked by 
the circles sampling the $>99\%$ convergence region in
Figure~\ref{fig:deconv-contour}. 
Figure \ref{fig:dcvlshapeerror} shows the measurement error
$\langle\Delta\eta_+\rangle$ for a circular PSF, and 
Figure~\ref{fig:dcvlshapebias} shows the shape bias induced by a PSF
ellipticity of $e_{\rm PSF}=0.1$.
The content of these figures are similar to those on the top and 
bottom row of Figure~\ref{fig:dcvlshapeerror100}, respectively.  
Evaluated at input $e=0$, the initial slope in $\langle\Delta\eta_+\rangle$
and the value of the PSF bias increase as $\sim1/\nu$ as $\nu$ decreases.
In both cases, there is a turn-around in the magnitude of the 
systematic, with a zero crossing in the region $0.6\,\simleq\,e$.
A rough estimate for the shear error $\Delta\gamma/\gamma$ due to 
these shape errors (assuming a flat $e$ distribution) is 
1--2\%.

Figure~\ref{fig:dcvlerrorestimate} plots the RMS error ratio of the 
actual to the estimated 
$\sqrt{\langle\Delta\eta_+^2/\tilde{\sigma}_{\eta_+}^2\rangle}$ 
for deconvolution fits, and is analogous to
Figure~\ref{fig:nativeerrorestimate}.  For the deconvolution,
the error estimate $\tilde{\sigma}_{\eta_+}$ does fairly well for $\nu\geq50$,
while $\sqrt{\langle\Delta\eta_+^2/\tilde{\sigma}_{\eta_+}^2\rangle}$ 
$\simeq1-\frac{2}{\nu}$ when $\nu<50$.

\subsubsection{The Airy Fix}

The open-circle points in the rightmost column of panels in
Figure~\ref{fig:dcvlshapeerror100} mark a tendency for the
deconvolution procedure to underestimate the ellipticity of large
objects.  This can be traced to the fact that the Airy function
is rather poorly described by a Gauss-Laguerre expansion.  In our
default procedure, the size of the basis set of the EGL expansion of
the PSF is made similar to the size of the PSF itself; in this case
the Airy function is poorly modelled at radii $\gg \lambda/D$.  For
example the Airy function has divergent second radial moment, while
any EGL expansion at finite order has a finite second radial moment,
indicating that the EGL expansion has failed to capture the large-$r$
behavior of the Airy function.
It takes GL order $N_{\rm GL}\equiv p+q=8$ and $16$ to 
describe the first and second Airy ring, respectively; our simulations
use GL order 12 to describe the PSF.

An equivalent statement is that, in the Fourier domain, an EGL
expansion fails to properly describe the small-$k$ behavior of the
Airy function.  This is not surprising since the Airy function has a
cusp at $k=0$ (which is equivalent to having infinite second moment in
real space).  The cusp in turn results from the sharp edge of the
illumination function of an unapodized circular mirror.

The poor description of the Airy function at large $r$ or small $k$
becomes important when trying to deconvolve images of well-resolved
galaxies, because the shape information for large galaxies is carried
at large $r$ and small $k$ relative to the PSF.  The finite EGL
expansion of the Airy PSF underestimates its circularizing effect on
large galaxies, hence the deconvolved shapes are too round for large
galaxies. Marginally-resolved galaxies don't have this difficulty because
they carry their ellipticity information in the part of $k$-space
where the EGL expansion is a good match to the Airy function.

We see that this ``Airy failure'' is not an intrinsic difficulty of
the shape-measurement methodology, but rather stems from a poor model
of the PSF.  This leads us to consider several possible solutions:
\begin{itemize}
\item Describe the PSF using a more appropriate function set than the
  Gauss-Laguerre expansion, if one expects nearly  diffraction-limited
  images.  This would complicate numerical calculation of the operator
  $C$ from \S\ref{fits}, but should fully restore the numerical accuracy for
  any PSF.
\item When fitting well-resolved galaxies, choose a basis for the EGL
  expansion of the PSF which matches the size of the galaxy rather
  than the size of the PSF.  This puts the fitting freedom of the EGL
  expansion in the range of $k$-space that is more important for the
  problem at hand; the core of the Airy function will be poorly fit
  but the wings will be better.
\item Change the PSF.
\end{itemize}
In these tests we implement the last of these three options, by
blurring the postage-stamp and PSF images with a Gaussian that is 
a $1/4$ of the size of the galaxy, and then
apply \glfit\ deconvolution using the smeared image pair.  
The blurring decreases the size mismatch between the PSF and galaxy GL
bases, which improves the accuracy of the GL expansion at the size
scale relevant to the large galaxy.  The blurring very slightly
increases the scatter of the shape measurement, but greatly reduces
the bias, as seen in the figure.

To summarize, the cusp at $k=0$ in the Airy PSF is not well fit by our
default PSF characterization, leading to biases in the shape
measurements.  This can be remedied in a number of ways without
invalidating our general approach.

\begin{figure}[!tbp]
\plotone{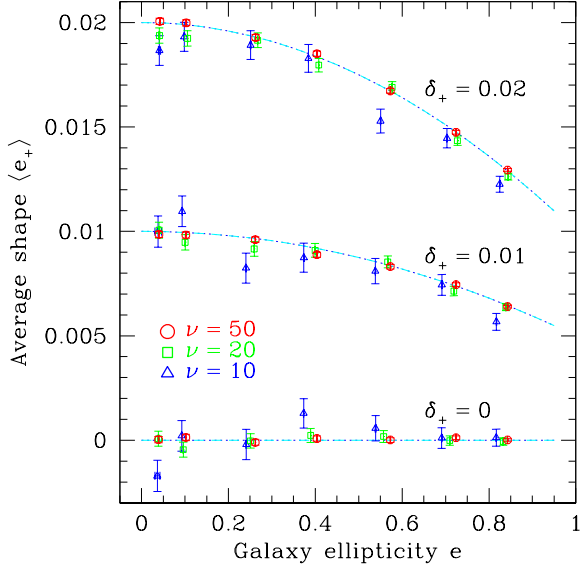}
\figcaption{Average shape $\langle e_+ \rangle$ obtained from 
a native fit ring test.
The horizontal axis plots the measured galaxy ellipticity $e$.  
Each point is an average from $10^5$ galaxies, while the error bar
represents the scatter in $\langle e_+\rangle$; the dot-dashed line 
indicates the expected signal ${\mathcal R}\delta_+$.
The $\langle e_\times \rangle$ response, not shown in this plot,
exhibit a similar response as for the case $\delta_+=0$, 
since $\delta_\times=0$.
\label{fig:nativeshear}}
\end{figure} 

\begin{figure}[!tbp]
\plotone{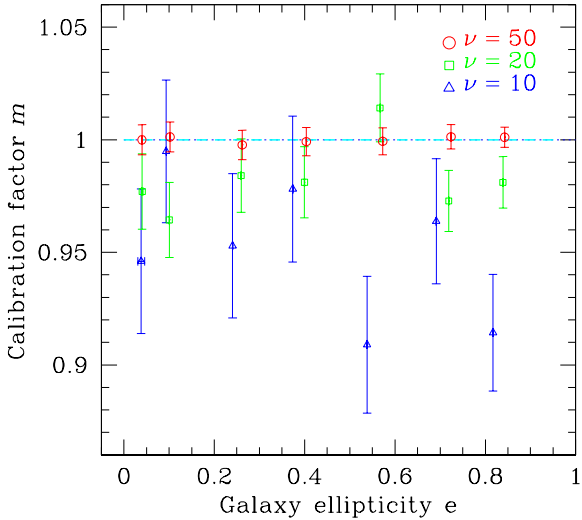}
\figcaption{Calibration factor $m$ in native fits (no PSF).
The vertical axis plots the normalized shear signal 
$\langle e_+ \rangle /({\mathcal R}\delta_+)$ from 
Figure~\ref{fig:nativeshear}
(the data for $\delta_+=0.01$ and $0.02$ have been combined).
The offset from unity is the calibration error.
\label{fig:nativeshearnorm}}
\end{figure}

\subsection{Accuracy for Asymmetric Galaxies}
\label{shearestimate}

The previous section discussed shape measurement accuracies with 
a well defined input shape and high degree of symmetry.  However, real
galaxy images 
do not have an unambiguous shape nor a perfect exponential
radial profile.

In this section, we measure the shapes of asymmetric galaxy images
that have irregular isophotes (\S\ref{imagemodel}), for which
there are no unambiguous definition of shape.
However, in weak lensing, it is not the accuracy of the individual 
shape measurement, but rather the ability to recover the shear from 
an ensemble of shapes, that is well-defined and needs to be accurate.  
In this sense, the ring test provides a measure of accuracy 
for assigning a shape for these irregular galaxies.

In the ring test, the shape $e$ is determined by \glfit\ measurements; 
the same image, before pixelization, is rotated by a series of
angles equally spaced between $0<\beta<2\pi$.
Each rotated image is then sheared by $\delta_+=0$, $0.01$ or $0.02$,
then pixelized for the ring test.  
With $100$ different orientations and $1,000$ (or more) realizations each, 
the uncertainty in the mean shape is 
$\sigma_{\delta_+}=\sigma_{e_+}/\sqrt{10^5}$,
where $\sigma_{e_+}$ is the scatter in the $e_+$ measurements
due to photon statistics.  The ring test has no intrinsic-shape noise.

We test both the native and deconvolution fit cases at a series of
different significance $\nu$, resolution $r_b$, and galaxy ellipticity $e$.
The asymmetric galaxies of different shape $e$ are distinct combinations 
of exponential-disk and deVaucouleur profiles; 
they are not sheared versions of the same underlying shape.

Additionally, the shear calibration factor $m$ and additive bias $c$
can be measured with the ring test.  The additive bias is the 
non-zero offset of the average shape when setting $\delta=0$ in 
the ring test, and is typically present when $e_{\rm PSF}\neq0$.  
The calibration factor $m$ is then obtained by repeating the ring test 
with $\delta\neq0$.

\subsubsection{Native Fits}
\label{nativeshear}

Figure~\ref{fig:nativeshear} shows the mean shape
$\langle e_+ \rangle$ as a function of the galaxy ellipticity $e$
for native fits.
The mean shape is expected to follow the curve ${\mathcal R}\delta_+$.
The approximation ${\mathcal R}=1-{e^2}/{2}$,
valid to ${\mathcal O}(\delta^3)$, is more than adequate here.
For the various significance ($\nu = 10$, $20$, or $50$),
the mean shape follows the expected curve very well.
In particular, we see that there is no significant 
additive bias $c$ seen in the $\delta_+=0$ data points,
to an uncertainty of $\sigma_c=\frac{0.01}{\nu}$.

Figure~\ref{fig:nativeshearnorm} plots the calibration factor $m$
by normalizing $\langle e_+ \rangle$ for the $\delta_+\neq0$ 
cases to the expected signal ${\mathcal R}\delta_+$,
$m = \langle e_+\rangle/({\mathcal R}\delta_+)$.
The calibration is well within 1\% of unity at $\nu=50$, 
but at lower detection significance, there is a systematic underestimation
of the order $m\simeq1-0.5/\nu$.

\subsubsection{Deconvolution Fits}
\label{deconvshear}

\begin{figure}[!tp]
\plotone{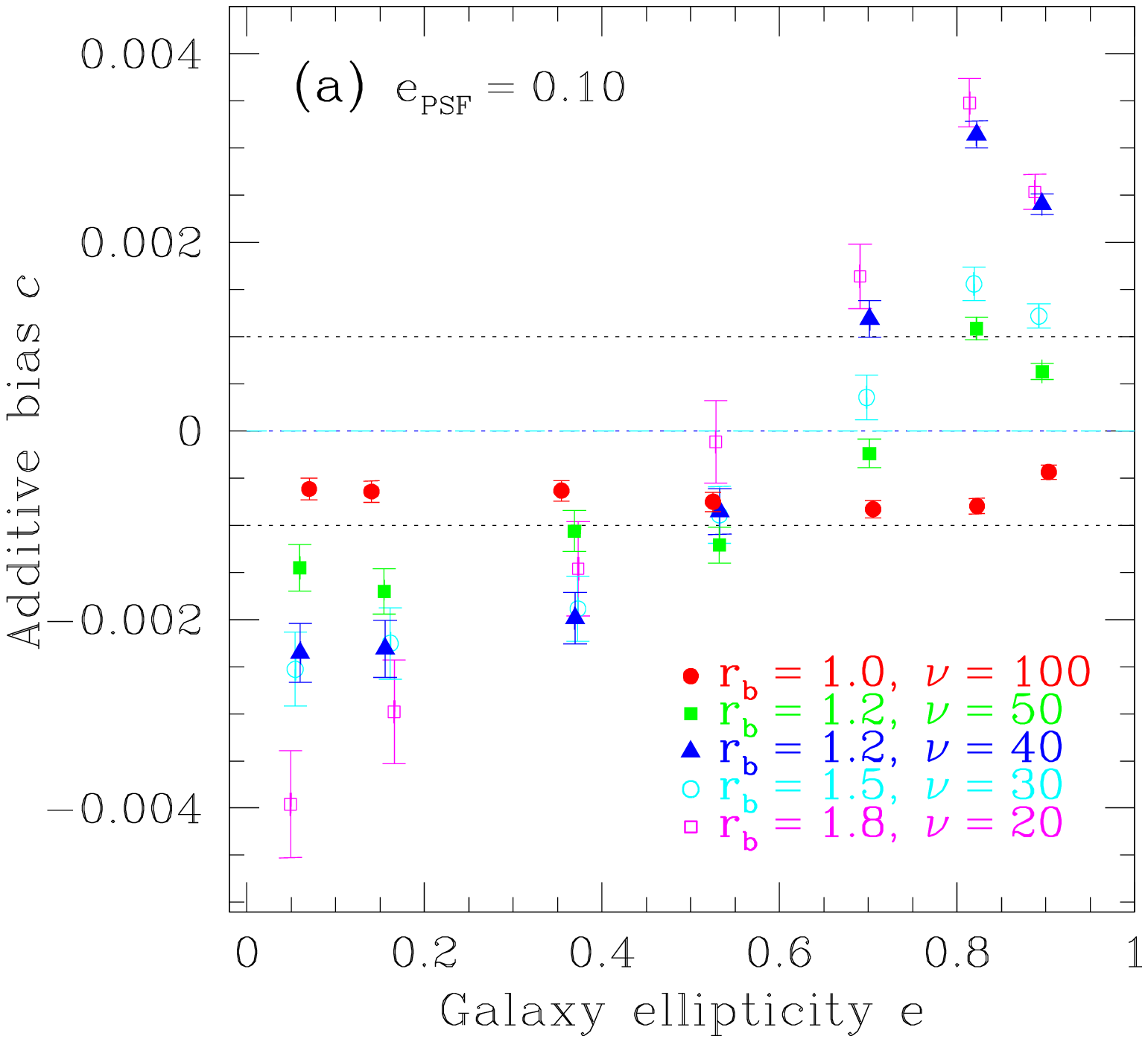}
\vspace{0.005in}
\plotone{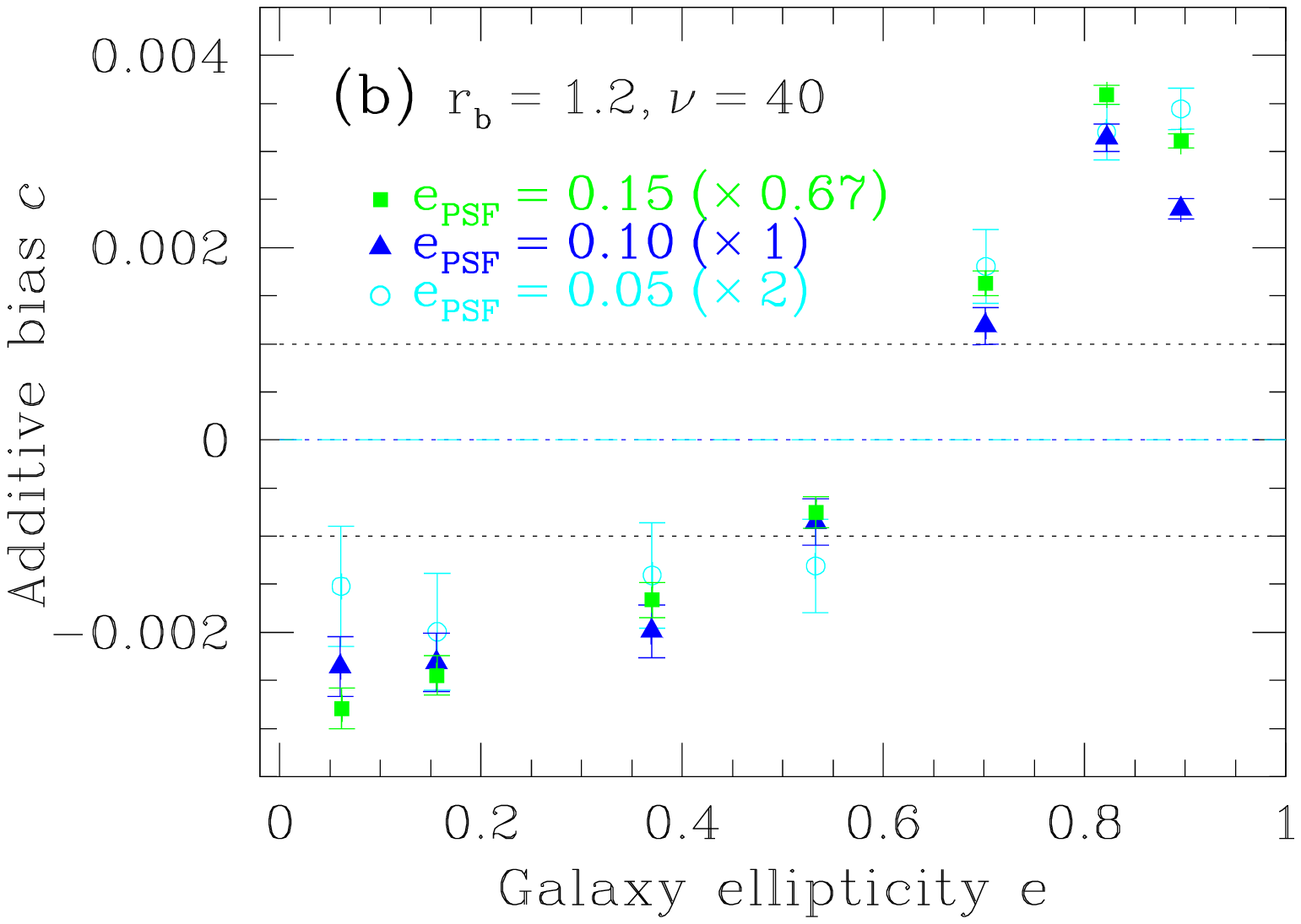}
\figcaption{Additive bias $c$ due to PSF anisotropy.  
The average shape $\langle e_+ \rangle$ is obtained from 
a deconvolution fit ring test with an Airy PSF of $e_{\rm PSF}=0.1$ 
for $\delta_+=0$, at various significance $\nu$ and resolution $r_b$.
The average shape is expected to lie at zero (dot-dashed line);
the offset from zero is the additive bias $c$.
The dotted lines indicate $\pm1\%$ of the PSF anisotropy. 
Each point is an average of $10^5$ galaxies.
(a) Additive bias at $e_{\rm PSF}=0.10$, for various $(\nu,r_b)$ pairs.
(b) Additive bias at $(r_b=1.2,\ \nu=40)$ for 
$e_{\rm PSF}=0.05, 0.10$ and $0.15$, with a scaling factor
normalized to $e_{\rm PSF}=0.10$.
\label{fig:dcvlshear}}
\end{figure} 

\begin{figure}[!tbp]
\plotone{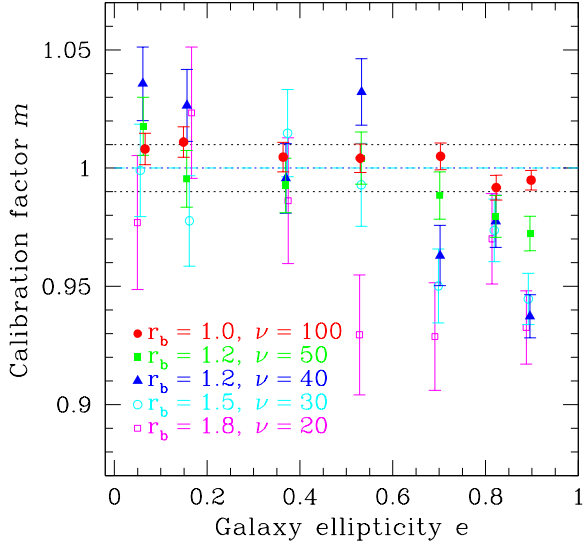}
\figcaption{Multiplicative bias in deconvolution fits, for various
$(\nu,r_b)$ pairs.  The shear is $\delta_+ = 0.02$, and  
the Airy PSF is circular, $e_{\rm PSF,+} = 0$.  
Each point is an average of $10^5$ galaxies.
The dotted line above and below indicate $\pm1\%$ of the shear.
({Currently running more simulations to decrease the error bar.})
\label{fig:dcvlshearnorm}}
\end{figure}

The ring test with asymmetric galaxies is similarly performed 
with deconvolution fits, with a circular Airy PSF of $e_{\rm PSF}=0$
and an elliptical PSF with $e_{\rm PSF}=0.1$.
Figure~\ref{fig:dcvlshear} shows the additive bias $c$ in
the shape average $\langle e_+\rangle$ when $\delta_+=0$ and
$e_{\rm PSF}=0.1$,
while Figure~\ref{fig:dcvlshearnorm} show the calibration factor $m$
for $\delta_+=0.02$.
Both Figures show various cases of $(r_b,\nu)$ pairs, which
traces the $99\%$ convergence contour from Figure~\ref{fig:deconv-contour}.

Figure~\ref{fig:dcvlshear}~(a) shows $c$ as a function
of $e$ at various $(r_b,\nu)$ pairs.
There is an $e$ dependence to the additive bias, which 
increases in magnitude with decreasing $\nu$.
Comparing with Figure~\ref{fig:dcvlshapebias}, we see that
the additive bias is essentially identical to the PSF bias found 
in symmetric galaxies, indicating that the PSF bias is independent 
of the galaxy symmetry, or the orientation of the galaxy shape 
over the ensemble average.
Figure~\ref{fig:dcvlshear}~(b) shows that for a given $(r_b,\nu)$,
the additive bias $c$ scales linearly with $e_{\rm PSF}$
for reasonable $e_{\rm PSF}$ values.

Figure~\ref{fig:dcvlshearnorm} shows the calibration factor $m$
for $e_{\rm PSF}=0$, $\delta_+=0.02$.
At $\nu=100$ and better, the measured shear is within 
$1\%$ of the input shear signal.
As the significance decreases ($\nu\leq50$), there is a
multiplicative bias for galaxy shapes over $e\,\simgeq\,0.6$,
whose magnitude is about $-2.5\%$ at $e=0.8$ (axis ratio $a/b=3$).
Similar results are found for $m$ for the case $e_{\rm PSF}\neq0$,
where $m = (\langle e_+\rangle - c)/ {\mathcal R}\delta_+$, but with
increased error bars due to the additional uncertainty in $c$.

\section{Discussion}

\subsection{Limitations of \glfit\ and Implication to WL Surveys}

There are limitations to \glfit\ with implications for WL surveys.
If \glfit\ is to be used for shape analysis, the sampling rate must be
such that  
the PSF minor-axis FWHM is sampled by $\ge2.8$ pixels; 
otherwise, \glfit\ fails to converge to a unique shape
(\S\ref{nativeconvergence}).  

Similarly, the PSF-smeared objects which produce useful shapes must have 
a minimum resolution that depends on the significance (or vice-versa)
in order to converge to a measurement of the pre-seeing galaxy shape.
If we require shape measurements of all values of $e$ to be successful, 
then $r_b\geq1.8$ is required at $\nu=20$, or
$r_b\geq1.2$ at $\nu\geq40$ (\S\ref{deconvconvergence}).
This implies that improvements in WL statistical accuracy are 
rapidly limited by the resolution of the galaxies:
although going deep in the exposure increases the $S/N$ of the
detected objects and reduces the scatter and systematic in the shear 
estimate, the required $S/N$ grows very rapidly for poorly resolved
images, and it becomes difficult to produce additional useful WL data
for poorly resolved galaxies.

We have also found that larger, well resolved galaxies
($r_b\simgeq\,5$) exhibit a problem with
the deconvolution method when the PSF is an Airy function, due to the
poor approximation of the Airy-functions wings by the GL expansion.
A simple remedy is available: smooth the PSF and galaxy images before
measuring both.  
Another solution is to choose an alternative to Gauss-Laguerre as 
the PSF decomposition basis functions, one that is better suited to describe 
an Airy function.  However, a non-GL decomposition will make the
deconvolution process excessively complex;  another solution 
is to apodize the telescope to suppress the Airy wings.

\subsection{Comparison to Other Methods}

\citet{Kuijken} offers an excellent description of the difference between
various WL techniques; we refer the reader to this paper for the difference
between, for example, the KSB method and Kuijken's method, which is
applicable to the difference bewteen KSB method and the EGL method.

Our EGL method is similar to Kuijken's polar shapelet method 
\citep{Kuijken}.  What the methods have in common are: 
the deconvolution of the PSF, which in principle allows for 
any PSF effects to be removed;
forward fitting, which allows error propagation, and hence 
an error estimate to the measured shape; 
and the definition of shape as shear, which has a well-defined 
shear transformation.
All of these features contribute to a better shear accuracy.

The differences between EGL and Kuijken's methods are subtle.
The first difference is that Kuijken method works the 
deconvolution and shearing in shapelet-coefficient space.
Our EGL method determines the shape-as-shear by iteratively fitting
within the pixel space.
Secondly, Kuijken's method obtains the shape using a shear transformation
valid to first-order in $e$, which can be off by up to 10\% at 
$e\simeq0.9$ ($g=0.6$).
Our method uses basis functions that are elliptical, {\it i.e.} the
shear transformation is valid to all orders,
which allows for the shape to be measured accurately for any $e$.
The third difference is that, in Kuijken's method, only the 
$m=0$ terms are used to describe the galaxy.  This allows for
less coefficients needed, and hence is efficient.
In comparison, our method obtains the full set of coefficients
to the specified order, which theoretically is more accurate
under shearing or deconvolution.
Another difference is in the choice of scale radius.
Kuijken's method uses 1.3 times the best-fit Gaussian $\sigma$ for 
the basis function scale radius, which apparently is optimal for the 
first-order method;  EGL uses the best-fit elliptical
Gaussian size, which gives the optimal sensitivity to small shear
(BJ02 \S3.1).

Unlike the KSB method, our EGL deconvolution method,
the \citet{Kuijken} polar shapelet and the \citet{K00} methods do not
rely on the approximation that the PSF is circular up to a small
linear ``smear.''
These methods are expected to do well on theoretical grounds; 
their differences \citep[at least with][]{Kuijken} seem to 
matter at the 1\% level.

\subsection{Comparison to Other Tests}

As mentioned earlier, the main difference between this paper and
previous end-to-end tests \citep{erben01,bacon01,STEP} is 
that we test the individual steps of the analysis.
The difference is mainly in the distribution of galaxies:
these end-to-end tests mimic the distribution to an actual WL field image,
while we test at specific S/N and galaxy size (resolution) sets.
By controlling the exact distribution of galaxy shapes, we also eliminate
shape noise in the shear estimate, and are able to quantify the effect
that noise or resolution/sampling has on the shear.

The main differences from other ``dissection'' methods 
\citep{Hirata03,Kuijken} are the inclusion of and quantifying noise effects,
testing with asymmetric galaxies, and determining the limits of
shape-measurement convergence.
\citet{Hirata03} demonstrates the calibration errors for 
a variety of PSFs (Gaussian, seeing-limited, diffraction-limited),
while we only test for Airy PSFs as the worst-case scenario.  
In \citet{Hirata03}, the galaxy types were varied as well 
(Gaussian, Exponential, deVaucouleurs), but they are symmetric, 
which can mask potential problems in shape measurements.
Their simulation focuses on the calibration accuracy, but does not 
quantify the additive error or include noise effects. 
\citet{Kuijken} includes an analysis with noise,
but does not quantify them; a variety of Sersic indices (for galaxies)
are tested, but are symmetric.
Their PSFs do not test the diffraction-limited (Airy) case,
although asymmetric PSFs, which are not considered in this paper, are
tested.

\subsection{Accuracy with EGL Method}

With the EGL method,  the systematic due to 
the shape measurement method is well under the 0.1\% level 
under ideal native-fit conditions, where the galaxy is 
symmetric and PSF is absent, as long as the minimum sampling 
and S/N criteria (minor axis width $>2.8$ pixels, $\nu>10$) are met.

The deviation from this high accuracy comes from real-life
complications.  In native fits, when the galaxies are not 
elliptically symmetric, the shear calibration factor degrades by 
$\sim-0.5/\nu$ as S/N is decreased.

The presence of a PSF further complicates the matter;
deconvolution with PSFs of truncated coefficients becomes a source of error
for well-resolved objects with an Airy PSF (up to 0.5\%), but
solutions to this poor approximation of the PSF are straightforward.
More generally, the circularly-symmetric Airy PSF itself introduces an
$e$ dependent 
systematic in the calibration factor, at worst $\sim-3\%$ at
$e=0.8,\;\nu=20$. 
A non-circular PSF adds a shear additive error that is 
proportional to the shape of the PSF, which also increases with 
decreasing S/N (up to $\pm3\%$ of $e_{\rm PSF}$ at $\nu=20$).
For the deconvolution fits, the sign and magnitude of the 
calibration and additive error is approximately the same 
for both the symmetric and asymmetric galaxies.

The systematics are typically inversely proportional to the S/N.
At high S/N ($\nu>50$), we have calibration error $\simleq1\%$, and 
the PSF anisotropy is $>99\%$ suppressed.  
These errors become $\sim4\%$ problems as $\nu\rightarrow20$.
The systematics are a function of the galaxy shape itself as well,
with the systematic at low $e$ having the reverse sign of that at 
high $e$.

To attain 0.1\% accuracy that is desired in future WL surveys, 
it is obvious that this shape measurement method itself needs to be refined.  
In addition, the whole pipeline needs to be rid of any systematics that are 
unrelated to the shape measurement, such as crowding or selection effects.

\section{Conclusion}

The elliptical Gauss-Laguerre decomposition is one of the most 
stringently tested methods to characterize shapes of galaxies.
With the EGL decomposition, shapes are measured without the need for
empirical shear/smear polarizabilities, and PSFs are removed by deconvolution.
The shear, obtained from averaging an isotropic ensemble of galaxy shapes,
is highly accurate due to the definition of shape as shear.

We have demonstrated that the EGL method allows shear recovery of
unprecedented accuracy, and quantified its degradation due to PSF, 
truncation of the EGL decomposition, image noise, and 
sampling~rate/resolution.  However, the current work is limited to the 
extraction of shapes; further work, including the full pipeline analysis,
is required for attaining 0.1\% accuracy in shear estimation.

\acknowledgements
We would like to thank David Rusin and
Mike Jarvis for much assistance and consultation.  This work is
supported by grants AST-0236702 from the 
National Science Foundation, Department of Energy grant
DOE-DE-FG02-95ER40893 and NASA BEFS-04-0014-0018.

\end{document}